\documentclass{aa}
\usepackage[varg]{txfonts}
\usepackage{hyperref}
\hypersetup{
  colorlinks=true,
  allcolors=[rgb]{0,0,0.8},
}

\begin{document}

\newcommand{\Healpix}{\texttt{HEALPix}\xspace}

\title{Direct measurement of 3D matter density profiles of galaxy clusters and a feature at the turnaround radius}

\author{%
  T.~Mistele\inst{1}\corrauth{mistele@mpe.mpg.de}
  \and E.~Bulbul\inst{1, 2}\email{ebulbul@mpe.mpg.de}
  \and M.~Kluge\inst{1}\email{mkluge@mpe.mpg.de}
  \and X.~Zhang\inst{1}\email{xzhang@mpe.mpg.de}
  \and S.~Grandis\inst{3}\email{sebastian.grandis@uibk.ac.at}
  \and E.~Artis\inst{1}\email{eartis@mpe.mpg.de}
  \and M.~Brüggen\inst{4}\email{marcus.brueggen@uni-hamburg.de}
  \and Z.~Ding\inst{1}\email{zding@mpe.mpg.de}
  \and L.~Fiorino\inst{1}\email{lfiorino@mpe.mpg.de}
  \and D.~Gruen\inst{5,6}\email{daniel.gruen@lmu.de}
  \and N.~Malavasi\inst{1}\email{malavasi@mpe.mpg.de}
  \and A.~Merloni\inst{1}\email{am@mpe.mpg.de}
  \and D.~Nagai\inst{7}\email{daisuke.nagai@yale.edu}
  \and K.~Nandra\inst{1}\email{knandra@mpe.mpg.de}
  \and M.~Ramos-Ceja\inst{1}\email{mramos@mpe.mpg.de}
  \and J.~Sanders\inst{1}\email{jsanders@mpe.mpg.de}
  \and A.~von der Linden\inst{8}\email{anja.vonderlinden@stonybrook.edu}
  \and S.~Zelmer\inst{1}\email{szelmer@mpe.mpg.de}
}

\institute{%
  Max Planck Institute for Extraterrestrial Physics, Gießenbachstraße 1, 85748 Garching, Germany
  \and
      Faculty of Physics, Fakult\"at f\"ur Physik, Ludwig-Maximilians-Universit\"at, Scheinerstr. 1, 81679 M\"unchen, Germany
  \and Universität Innsbruck, Institut für Astro- und Teilchenphysik, Technikerstr. 25/8, 6020 Innsbruck, Austria
  \and Universität Hamburg, Hamburger Sternwarte, Gojenbergsweg 112, 21029 Hamburg, Germany
  \and University Observatory, Faculty of Physics, Ludwig-Maximilians-Universität, Scheinerstr. 1, 81679 Munich, Germany
  \and Excellence Cluster ORIGINS, Boltzmannstr. 2, 85748 Garching, Germany
  \and Department of Physics, Yale University, New Haven, CT 06520, USA
  \and Department of Physics and Astronomy, Stony Brook University, Stony Brook, NY 11794, USA
}

\date{\today}

\abstract{%
The infall regions of galaxy clusters offer a unique window into the transition between virialized halos and the surrounding large-scale structure, serving as a probe of the growth of structure, matter accretion, and cosmology. We present the first direct, non-parametric measurements of the spherically averaged three-dimensional density profiles of galaxy clusters extending deep into their infall regions, out to $35\,\mathrm{Mpc}$, with significant constraining power. We infer weak-lensing shear profiles from the DES Year 3 and DECADE shape catalogs for the clusters detected in the deepest SRG/eROSITA Survey and apply a non-parametric deprojection technique to reconstruct the spherically averaged density profiles. We independently measure the profiles for clusters in the DES and DECADE footprints, providing two statistically independent measurements. Both show an $r^{-3}$ density decline over an extended radial range in the one-halo regime, in agreement with expectations in the $\Lambda$CDM model. At larger radii, however, we find a distinct non-monotonic feature near the expected location of the turnaround radius. The feature is independently detected in both DES and DECADE and is statistically significant, with a simple broken power-law description ruled out at more than $5\sigma$, and is robust to a range of potential systematic effects tested in this work. Its physical origin remains unclear, but its location and its scaling with cluster size suggest a connection to the dynamics of the outer infall region and, potentially, to the turnaround radius. These results demonstrate the power of non-parametric density reconstructions to reveal features in the outskirts of clusters that are not captured by simple parametric models, opening a new observational window into the connection between galaxy clusters, their surrounding large-scale structure, and cosmology.
}


\maketitle
\nolinenumbers %

\section{Introduction}

The mass distribution in the outskirts of galaxy clusters, extending beyond the virial radius, encodes information on the nature of dark matter, the physics of structure formation, and the underlying cosmological model \citep{Diemer2015, Adhikari2014,Mpetha2024}. While the abundance and total masses of galaxy clusters provide powerful constraints on cosmology, as well as the parameter space of well-motivated dark matter models \citep[e.g.][]{Artis2024, Artis2025, Zelmer2025, Strunk2026}, spatially resolved mass profiles provide complementary information by probing the precise radial distribution of matter and the transition from the virialized halo to its surrounding large-scale structure.

Within the $\Lambda$CDM framework, the matter distribution around halos can be described in terms of a one-halo contribution, dominated by orbiting matter gravitationally bound to the cluster, and a two-halo contribution from correlated large-scale structure in the infall region \citep{Gill2005}.  
Beyond the virialized region, the transition from collapsed to infalling material changes the slope of the cluster mass density profile, associated with the splashback feature \citep{Adhikari2014, More2015}, which approximately marks the apocenters of recently accreted material and provides a physically motivated boundary of the halo. The location and sharpness of this feature depend on halo mass, redshift, and mass-accretion history, thereby providing information on halo growth and the dynamics of structure formation \citep{Diemer2014}. The extended infall region, where clusters continue to accrete matter from the surrounding large-scale structure beyond the virial radius, typically spans $\sim1$–$5$~Mpc from the cluster center. Its outer boundary is set by the turnaround radius, which separates infalling matter from matter receding with the cosmic expansion \citep{Gunn1972, Pavlidou2014, Fong2021}. Because this regime is sufficiently far from the dense central environment, the matter distribution is expected to be comparatively insensitive to baryonic feedback processes \citep[e.g.][]{Bahar2024}, making it a particularly clean tracer of the cluster's assembly, growth history, and cosmology \citep{Mpetha2024, Mpetha2025}.

Recent advances in wide-area X-ray, Sunyaev-Zel'dovich Effect, and weak-lensing surveys have opened new avenues for probing the matter distribution and the physical processes governing matter accretion and gravity in the infall regions of galaxy clusters. Until recently, the extremely low X-ray surface brightness of these regions made them largely inaccessible to direct observation. The advent of the extended ROentgen Survey (eROSITA) All-Sky Survey, combined with stacking analyses of large cluster and group samples, has enabled the detection of X-ray emission out to several virial radii ($\sim4.5$~Mpc), revealing shock-heated gas in the accretion region and a transition from virialized to infalling gas beyond the virial radius in galaxy clusters and groups \citep{Zhang2024, Zhang2025, Bulbul2026}. These results complement recent observations of the baryonic component of cluster outskirts using the thermal Sunyaev--Zel'dovich effect \citep{Anbajagane2022, Anbajagane2024, Cheng2026}, and X-ray emission \citep[e.g.,][]{Zhang2025}. Together, these probes provide a multi-component view of the matter distribution in the infall region, linking the gas thermodynamic state to the underlying dark-matter-dominated mass distribution traced by weak lensing. 

Advances in wide-field imaging surveys and weak lensing analyses have enabled measurements of the transition from the virialized halo into the infall regions. Weak gravitational lensing provides a direct probe of the projected mass distribution around galaxy clusters, independent of the dynamical state of the matter \citep[e.g.,][]{Bartelmann2001, Umetsu2020}. In this way, weak-lensing measurements provide observational evidence for, for example, the splashback radius \citep{Umetsu2017, Chang2018, Contigiani2019}, first identified through the spatial distribution of galaxies around clusters \citep{More2016, Rana2023}, demonstrating that lensing can trace the cluster mass distribution beyond the overdensity radii, R$_{500}$, and into the infall regions of clusters.

In current weak lensing analyses, however, the underlying density profiles are often modeled by assuming a parametric form, and its predicted projected shear is forward-modeled and fitted to the data. While this approach provides robust mass estimates, it can impose assumptions on the shape of the density profile precisely in the regime where departures from a simple halo description are physically informative. An alternative method removes this limitation by reconstructing the mass distribution directly from the observed lensing signal without imposing a parametric form for the density profile. Such non-parametric or direct deprojection approaches have recently been applied to cluster weak-lensing data \citep{Mistele2023d, Mistele2024, Mistele2024b, Umetsu2025}, with a particular focus on the enclosed mass rather than directly recovering the density profile. 
A direct measurement of the density profile provides a complementary view of the transition from the virialized halo through the infall region and toward the turnaround region. A deprojection method that achieves this was developed and employed by \citet{Mistele2025}.
It was successfully applied to the CLASH sample of galaxy clusters \citep{Postman2012} out to relatively small cluster-centric radii, and the available sample size did not provide sufficient statistical power to robustly characterize the density profile deep into the outskirts.
A non-parametric density deprojection was first employed by \citet{Johnston2007b}, who studied optically selected SDSS clusters, but based all their conclusions on mass profiles and parametric fits to shear profiles instead.

In this work, we present the first direct, non-parametric measurements of the three-dimensional density profiles of galaxy clusters extending into their infall regions and out to $35\,\mathrm{Mpc}$, with significant constraining power. Specifically, we measure the spherically averaged three-dimensional density profiles. For this, we utilize the largest intracluster medium selected cluster sample to date, detected in the 4.4 consecutive surveys carried out by the eROSITA X-ray Telescope on board the Spectrum--Roentgen--Gamma (SRG) mission \citep{Predehl2021, Sunyaev2021, Merloni2024} in the Western Galactic Hemisphere \citep[eRASS:5, hereafter,][]{erass5catalog}. We combine this cluster sample with wide-field weak-lensing measurements from the Dark Energy Survey Year 3 \citep[DES Y3,][]{Gatti2021} and the Dark Energy Camera All-Data Everywhere \citep[DECADE,][]{Anbajagane2025} data sets. This combination enables measurements of cluster density profiles with substantially improved statistical precision, allowing us to probe the transition from the virialized halo through the splashback region and into the infall regime. This allows us to, for the first time, identify a characteristic feature associated with the transition toward the turnaround region.
This paper is organized as follows. In Sect.~\ref{sec:data}, we introduce the cluster and weak-lensing data sets used in our analysis. In Sect.~\ref{sec:method}, we describe the method used to reconstruct the three-dimensional density profiles. We present the observational results in Sect.~\ref{sec:results} and validate our deprojection method using hydrodynamical simulations in Sect.~\ref{sec:method:hydro}. We carry out a number of robustness checks in Sect.~\ref{sec:results:checks}. We discuss the physical implications of our results in Sect.~\ref{sec:discussion} before summarizing our conclusions in Sect.~\ref{sec:conclusion}. Unless otherwise stated, we assume a flat FLRW background cosmology with $H_0 = 70\,\mathrm{km}\,\mathrm{s}^{-1}\,\mathrm{Mpc}^{-1}$, $\Omega_m = 0.3$, and $\Omega_r = 0$. Throughout the paper, projected (2D) radii are denoted by $R$, while deprojected (3D) radii are denoted by $r$.

\begin{figure}
  \includegraphics[width=\columnwidth]{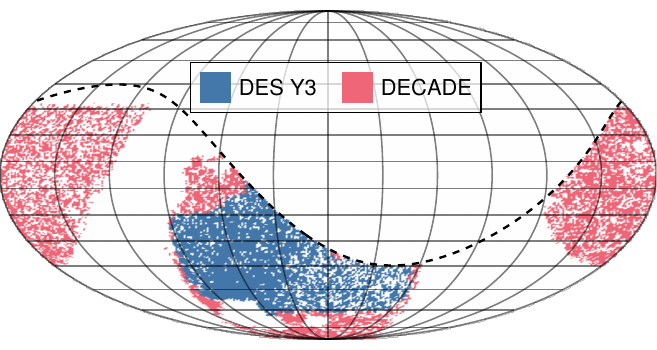}
  \caption{%
        Regions within the DES~Y3 (blue) and DECADE (red) footprint that contain at least one cluster from the eRASS:5 cosmology sample.
        We show a \Healpix pixelization with `NSIDE=64'.
        The dashed black line delineates the Western and Eastern Galactic hemispheres.
        Adapted from \citet{Mistele2026}.
    }
  \label{fig:footprints}
\end{figure}

\section{Data}
\label{sec:data}
In this section, we discuss the lens, i.e., cluster, and source galaxy samples used in our measurements. We utilize the galaxy clusters identified in the eRASS:5 cosmology sample from the Western Galactic Hemisphere \citep{erass5catalog}. The sample comprises 21,381 clusters spanning a redshift range of $0.1 \leq z_l < 1.1$. The clusters are X-ray selected using an extent-likelihood threshold of $\mathcal{L}_\mathrm{ext} > 6$ \citep[e.g.][]{Bulbul2024} and are optically identified with the eROMaPPer cluster-finding pipeline \citep{Kluge2024}. We further apply a $\lambda \geq 15$ cut in optical richness in order to minimize contamination, and we restrict the sample to $\lambda < 100$ in order to avoid high-richness clusters missing weak-lensing data in the outermost radial bins.
The latter is because, in our analysis below, we re-scale cluster-centric distances with cluster size, meaning high-richness clusters need shear out to larger distances, whereas, to speed up the analysis, we have pre-computed the list of lensing sources around each cluster out to a fixed $35\,\mathrm{Mpc}$.
We have verified that, nevertheless, including $\lambda \geq 100$ clusters does not significantly change the results.

Following \citet{Mistele2026}, we use the DES~Y3 \citep{Gatti2021} and DECADE \citep{Anbajagane2025} shape catalogs to measure the weak-lensing shear around our eRASS:5 cluster sample. Together, DES~Y3 and DECADE cover approximately 90\% of the clusters in our sample, with no overlap between their footprints. The spatial overlap between the eRASS:5 clusters and the DES~Y3 and DECADE footprints is shown in Fig.~\ref{fig:footprints}. Although the DECADE catalog has a somewhat lower effective number density than DES~Y3, its larger footprint compensates for this difference, resulting in comparable overall statistical power of the two datasets.
Our final sample contains 6,814 clusters in an area of 3,272 deg$^2$ overlapping with the DES Y3 footprint and 8,331 clusters in an area of 5,987 deg$^2$ overlapping with the DECADE footprint.
The signal-to-noise ratio in the stacked shear profiles of these clusters is 105 for DES Y3 and 107 for DECADE.\footnote{%
 We calculate these ratios as in \citet{Grandis2024} but we include all tomographic bins, do not split into redshift and richness bins, and do not include any factors of the critical surface density in the weights.
}

\section{Method}
\label{sec:method}
In this section, we describe how we measure the radial density profiles using weak-lensing data.
Before going into the details, we give a brief high-level overview.

\begin{figure}
  \includegraphics[width=\columnwidth]{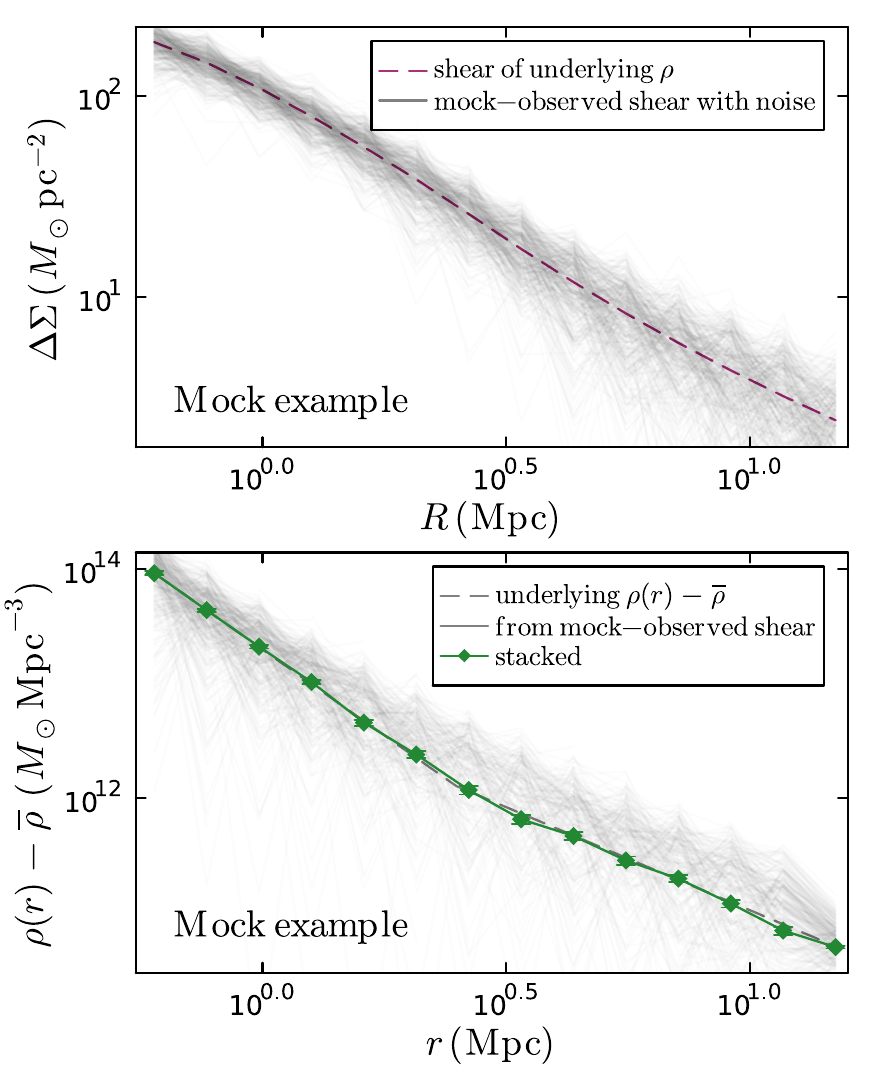}
  \caption{%
   Illustration of our shear-to-density conversion (i.e., deprojection) using a simple mock example.
   As the ``true'' underlying density profile we adopt a spherically symmetric broken power law with inner slope $-3$, outer slope $-1.8$, and a transition at $r = 2.5\,\mathrm{Mpc}$.
   At small radii, $r < 0.3\,\mathrm{Mpc}$, we take the density to be constant.
   \emph{Top}: The shear calculated from the underlying density profile (dashed purple line) and 400 mock observations (solid gray lines) including Gaussian noise.
   \emph{Bottom}: The noisy density profiles obtained by applying Eq.~\eqref{eq:rho_from_shear_demo} to each of the 400 noisy shear profiles (solid gray lines) and the density profile obtained by stacking these (green markers and green solid line).
   The noise cancels out when stacking and the underlying density profile (dashed gray line) is recovered within the uncertainties.
   }
  \label{fig:mocks}
\end{figure}

Our first step is to measure the shear profile of each cluster and to convert it to an excess surface density profile, $\Delta \Sigma$.
This conversion amounts to multiplying the shear by a geometrical factor called the critical surface density $\Sigma_{\mathrm{crit}}$ (see below for the definition of $\Sigma_{\mathrm{crit}}$).
A cluster's $\Delta \Sigma$ encodes information about its projected/2D mass distribution.
Mock-observed $\Delta \Sigma$ profiles are illustrated by the gray lines in the top panel in Fig.~\ref{fig:mocks}.

The second step is to convert the $\Delta \Sigma$ of each cluster to a deprojected/3D density profile.
This is where we go beyond traditional weak lensing analyses:
Instead of fitting a parametric profile \citep[such as an NFW profile,][]{Navarro1996} to the observed $\Delta \Sigma$, we directly and non-parametrically convert excess surface density $\Delta \Sigma$ to density $\rho$.
To this end, we use the deprojection formula from \citet{Mistele2025,Mistele2024b}, namely
\begin{multline}
 \label{eq:rho_from_shear_demo}
 \rho(r) - \overline{\rho} =
  \frac{1}{\pi r} \int_0^{\pi/2} d\theta \, \frac{\Delta \Sigma \left(\frac{r}{\sin \theta}\right) - \Delta \Sigma(r)}{\cos^2 \theta}
  \\
  + \frac{2}{\pi r} \int_0^{\pi/2} d\theta \, \Delta \Sigma\left(\frac{r}{\sin \theta}\right) \,,
\end{multline}
with the cosmological matter density $\overline{\rho}$.
Specifically, we obtain the overdensity $\rho - \overline{\rho}$, since weak lensing is not sensitive to a constant density.
This is illustrated by the gray lines in the bottom panel in Fig.~\ref{fig:mocks}.

Finally, we stack the overdensity profiles of all clusters by taking a weighted mean.
The profiles of the individual clusters are typically very noisy.
However, this noise cancels out if we stack sufficiently many clusters, since the relation between the observed $\Delta \Sigma$ and the inferred $\rho$ is linear.
This is illustrated by the green symbols in the bottom panel in Fig.~\ref{fig:mocks}.

\subsection{Shear profiles}
\label{sec:method:shear}

As mentioned above, our first step is to measure the azimuthally averaged tangential shear for each individual galaxy cluster as a function of projected radius $R$.
In particular, we are interested in the combination $\Sigma_\mathrm{crit} g_t$ where $g_t$ is the tangential reduced shear and $\Sigma_\mathrm{crit}$ is the critical surface density,
\begin{align}
 \Sigma^{-1}_{\mathrm{crit},ls} = \frac{4 \pi G_N}{c^2} \frac{D(z_l)}{D(z_s)} \cdot \max \{ 0, D(z_l, z_s) \}\,, 
\end{align}
with Newton's gravitational constant $G_N$ and the angular diameter distances to the lens, $D(z_l)$, to the source, $D(z_s)$, and between the source and the lens $D(z_l, z_s)$.
In the following, we denote the azimuthal average $\langle \Sigma_\mathrm{crit} g_t \rangle$ as $G$.

For an individual cluster $l$ and for a (projected) radial bin denoted by $R_{l i}$, we use the following standard estimator for $G$,\footnote{
 We first subtract the randoms' shear $G_\mathrm{random}$ and then apply the boost factor $B$ \citep[contra, e.g.,][]{Prat2022,Giannini2026}.
 This is the correct order of operations because, in contrast to the true shear signal, any additive shear bias is not diluted due to cluster member galaxies.
}
\begin{align}
 \label{eq:G_l}
 G_l (R_{l i}) = B_{l i} \cdot \left(
   \frac{
    \sum_{s \in {l i}} W_{ls} \cdot 
    \Sigma_{\mathrm{crit},ls} \mathcal{R}_s^{-1} \epsilon_{t, ls}
   }{
    \sum_{s \in {l i}} W_{ls}
   }
   - G_\mathrm{random}^{l i}
 \right)\,,
\end{align}
where the weights $W_{ls}$ are given by
\begin{align}
W_{l s} = \Sigma_{\mathrm{crit},ls}^{-2} \mathcal{R}_s w_s\,.
\end{align}
Here, $w_s$ is the weight of the source $s$, $\epsilon_{t,ls}$ is the tangential ellipticity of the source $s$ with respect to the location of the lens $l$, and $\mathcal{R}_s$ is the shear response of the source.
The notation ``$s \in {l i}$`` indicates that the sum runs only over sources that are located within the given radial bin $i$ around lens $l$.
The boost factor $B_{li}$ and the randoms' shear profile $G_\mathrm{random}^{li}$ are discussed separately below.
We do not include the selection response here.
Further, following \citet{Grandis2024}, we smooth the shear response using a 2D signal-to-noise and size-ratio grid \citep[see also][]{Gatti2021,Anbajagane2025}.
As we will show explicitly below, including the selection response and using the non-smoothed shear response do not significantly change our results.

We take $\Sigma_{\mathrm{crit},ls}$ to be $\langle \Sigma_{\mathrm{crit}}^{-1} \rangle^{-1}$, where the average runs over source redshifts weighted according to the calibrated `SOMPZ' redshift distributions \citep{Myles2021,Anbajagane2025b}.
For a given source, we use the `SOMPZ' of that source's tomographic bin.
These redshift distributions are provided by DES~Y3 and DECADE and, following \citet{Mistele2026}, we take them to be normalized to $1+m$, i.e., normalized to the multiplicative shear bias.
This multiplicative shear bias is then automatically corrected for through $\Sigma_{\mathrm{crit},l s}$.
A caveat is that these global `SOMPZ' redshift distributions are not matched to our choice of weights and our lens selection.
Below, we explicitly show that this has only a negligible effect on our results.

The sums over sources $s$ in Eq.~\eqref{eq:G_l} run over all sources, including foreground galaxies and cluster members.
Both of these dilute the signal since only background sources are actually sheared.
The presence of the foreground sources is taken into account by the `SOMPZ' redshift distributions used to calculate $\langle \Sigma_{\mathrm{crit}}^{-1} \rangle$.
In particular, we do \emph{not} re-normalize them to $1+m$ in the interval $[z_l, \infty)$ (the redshift range of background sources).
As a result, $\langle \Sigma_\mathrm{crit}^{-1} \rangle$ is implicitly multiplied by its probability of being in the background, which boosts $G$ accordingly.

Contamination from cluster member galaxies is taken into account through the boost factor $B_{li}$,
\begin{align}
 B_{l i} = \frac{
  \sum_{s \in {li}} W_{l s}
}{
  \sum_{s \in {li}} W_{l s} \left(1 - f_\mathrm{cl}^{l i s}\right)
} \,,
\end{align}
where the cluster member fraction $f_\mathrm{cl}$ is obtained from fits to the radial number density profiles as described in \citet{Mistele2026}. 
In particular, we fit the parametric form proposed in \citet{Grandis2024} in bins of richness and lens redshift.
This procedure may somewhat underestimate the extent of contamination due to obscuration and blending \citep{Kleinebreil2024,Schrabback2026}.
However, as we will see below, cluster-member contamination is not important for our results, and setting $B = 1$ does not significantly change them.
The $G_\mathrm{random}^\mathrm{li}$ in Eq.~\eqref{eq:G_l} represents the shear from random coordinates.
We subtract it to remove residual additive shear biases.
We explain our procedure in more detail in Sect.~\ref{sec:method:randoms}.
As we will see below, subtracting randoms is not important for our results, i.e., setting $G_\mathrm{random} = 0$ does not change our results significantly.

In this way, we obtain the shear profile $G_l (R_{l i})$ for each lens $l$ in discrete radial bins $R_{l i}$.
The index $l$ in $R_{li}$ is required because we often use different radial bins for different clusters.
This is because we rescale cluster-centric distances by (a proxy of) cluster size.
We take the value of $G_l(R_{l i})$ to be measured at the location $R_{l i}$ given by the weighted average of the projected distances $R_{l s}$ between the sources $s$ and the lens $l$,
\begin{align}
 \label{eq:R_li}
 R_{l i } = \frac{
  \sum_{s \in {l i}} W_{l s} (1 - f_\mathrm{cl}^{l i s}) \cdot R_{l s}
}{
  \sum_{s \in {l i}} W_{l s} (1 - f_\mathrm{cl}^{l i s})
} \,.
\end{align}
The factors of $1 - f_\mathrm{cl}$ ensure that cluster member contaminants, which do not contribute to the shear, also do not contribute to $R_{l i}$.

In our analysis below, we employ two different types of radial shear profiles: One uses radial bins in the standard way, i.e. the radial bins do not overlap and the shear in bin $i$ is measured from sources that are disjoint from the sources used in other bins $i'$.
The second measures the shear profile in overlapping sliding windows.
All formulas shown above are valid in both cases; the main difference is that, in the sliding window case, neighboring data points in the shear profile $G_l (R_{l i})$ are highly correlated.

\subsection{Converting shear to deprojected/3D density}
\label{sec:method:shear_to_rho}

Our second step is to convert the shear profile $G_l$ of each individual cluster $l$ to its 3D density profile $\rho_l$.
To this end, we use the non-parametric deprojection method from \citet{Mistele2025} \citep[which builds on earlier work in][]{Mistele2023d,Mistele2024,Mistele2024b}.
This method does not fit a parametric profile to the observed shear; instead, it directly converts the observed shear to density.

Specifically, we first convert the observed shear profile $G (R)$ to the excess surface density profile $\Delta \Sigma (R)$.
The difference between $G = \langle g_t \, \Sigma_\mathrm{crit} \rangle$ and $\Delta \Sigma$ goes to zero as $\Sigma/\Sigma_\mathrm{crit}$ goes to zero.
At the large radii we are interested in, $\Sigma/\Sigma_\mathrm{crit}$ is typically small.
Thus, we expect that the difference between $G$ and $\Delta \Sigma$ is not important for our results and we assume
\begin{align}
 G(R) \approx \Delta \Sigma (R) \,.
\end{align}
This is equivalent to assuming that shear and reduced shear do not differ.
Below, we will verify this assumption by re-running our analysis using the full $G$-to-$\Delta \Sigma$ conversion formula from \citet{Mistele2024b}.
We then convert $\Delta \Sigma (R)$ to $\rho(r)$ using the following deprojection formula,
\begin{equation}
 \label{eq:rho_from_ESD}
 \rho(r) - \overline{\rho} = \frac{I_1(r) + I_2(r)}{\pi r} \,,
\end{equation}
with
\begin{align}
 I_1(r) &= \int_0^{\pi/2} d\theta \,
  \frac{\Delta \Sigma \left(\frac{r}{\sin \theta}\right) - \Delta \Sigma(r)}{\cos^2 \theta} \,, \\
 I_2(r) &= 2 \int_0^{\pi/2} d\theta \, \Delta \Sigma\left(\frac{r}{\sin \theta}\right) \,.
\end{align}
As mentioned above, weak lensing is insensitive to any addititive constant density.
Thus, the result is $\rho(r) - \overline{\rho}$ and not $\rho(r)$, where $\overline{\rho}$ is the cosmological background matter density at the lens redshift $z_l$.\footnote{
 This was not made explicit in \citet{Mistele2024b,Mistele2025} because the contribution from $\overline{\rho}$ is negligible at the radii considered there.
 This is no longer the case in our present analysis.
}
Below, we will often consider the overdensity $\delta(r) = (\rho(r) - \overline{\rho})/\overline{\rho}$ that we obtain by calculating $\overline{\rho}$ in our assumed cosmology at the lens redshift and dividing Eq.~\eqref{eq:rho_from_ESD} by $\overline{\rho}$.

Mathematically, Eq.~\eqref{eq:rho_from_ESD} is equivalent to taking the derivative of the following deprojection formula for the mass profile $M(r)$ \citep{Mistele2024b},
\begin{align}
 \label{eq:M_from_ESD}
 M(r) - \frac{4 \pi}{3} \overline{\rho}_l r^3 = 4r^2 \int_0^{\pi/2} d\theta\, \Delta \Sigma\left(\frac{r}{\sin \theta}\right) \,,
\end{align}
and then dividing by $4 \pi r^2$.
The advantage of Eq.~\eqref{eq:rho_from_ESD} is that it does not require taking any numerical derivatives.
Of course, the finite difference $\Delta \Sigma(r/\sin \theta) - \Delta \Sigma(r)$ in the integral $I_1$ is still sensitive to noise, especially since the $1/\cos^2 \theta$ factor up-weighs precisely the region around $\theta \approx \pi/2$ where the difference becomes small.
Importantly, however, these formulas are linear in the observed shear $G$, so any noise will cancel out when stacking sufficiently many lenses (see Sect.~\ref{sec:method:stacking}).

In practice, we measure the shear profile $G_l$ of a lens $l$ only at discrete radii $R_{l i}$, while evaluating the integrals in Eq.~\eqref{eq:rho_from_ESD} requires $G_l$ at \emph{all} radii $R \geq r$.
Thus, we interpolate using quadratic B-spline interpolation.
We interpolate $G_l$ as a function of $\log_{10} R$.
We use quadratic rather than linear interpolation because tests with mock data show that this is less likely to wash out localized features.
The difference between linear and quadratic interpolation is insignificant when the value of $G_l$ is known on a fine radial grid, which can, for example, be achieved by using sliding windows rather than traditional radial bins to measure $G_l$.
Below, we show explicitly that the choice of linear vs quadratic interpolation does not affect our conclusions.
We also note that quadratic B-spline interpolation is still linear in the observed shear, so the noise still cancels out after stacking.
We similarly need to extrapolate $G_l$ beyond the last measured data point.
This is because the integrals in Eq.~\eqref{eq:rho_from_ESD} extend all the way to $R \to \infty$ (consider $r/\sin \theta$ for $\theta \to 0$).
In practice, most of the integral weight is close to $R \approx r$, and the extrapolation matters only when $r$ is close to the outermost shear data point.
We extrapolate assuming $G$ declines as $R^{-1}$.
Below, we show explicitly that this choice does not affect our conclusions.

We apply the above deprojection formulas to each individual cluster $l$.
The results of this procedure are the overdensity profiles $\rho_l (r) - \overline{\rho}_l$ of each cluster.
We evaluate these profiles at a set of discrete 3D radii $r_{l i}$, which we take to be the same as the projected radii $R_{l i}$ where we measured the shear profile.

The deprojection formulas discussed here assume spherical symmetry \citep[see the derivations in ][]{Mistele2024b,Mistele2025}.
However, as shown in \citet{Mistele2025}, the results are still meaningful outside spherical symmetry.
Indeed, if the clusters are randomly oriented relative to our line of sight, our deprojection formula does, on average, reproduce the true spherically averaged density profile even for a non-symmetric lens.
This is a consequence of the fact that the formulas are linear in the observed shear.\footnote{
 And we expect that taking into account the difference between shear and reduced shear, which makes the conversion from $G$ to $\Delta \Sigma$ non-linear, changes this only to a minor degree and only at small radii, see Fig.~4 in \citet{Mistele2025}.
}
Thus, if we stack sufficiently many lenses, the resulting stacked density profile matches the true spherically averaged density profile even for non-symmetric lenses.
This is explicitly illustrated using shear profiles of simulated non-symmetric mass distributions in Sect.~\ref{sec:method:hydro} below.
The assumption of random orientations between clusters and our line of sight is reasonable for X-ray-selected samples such as the eRASS:5 sample.
A secondary projection effect may, however, be introduced when applying additional richness-based cuts \citep[e.g.,][]{Dietrich2014,Abbott2020,Wu2022,Sunayama2020}, as we do below.
This is further discussed in Sect.~\ref{sec:results:selection} and Sect.~\ref{sec:method:hydro}.

\subsection{Stacking}
\label{sec:method:stacking}

Our third and final step is to stack the density profiles of the individual clusters by taking a weighted mean across all the clusters $l$ at the same radial bin $i$,
\begin{align}
 \rho_i = \sum_l w_{l i} \, \rho_l (r_{l i}) \,,
\end{align}
where the weights $w_{l i}$ are normalized, i.e. $\sum_l w_{l i} = 1$.
As is common in weak lensing analyses, we choose inverse variance weights,
\begin{align}
 w_{l i} \propto \sigma^{-2}_{\rho_l (r_{l i})} \,.
\end{align}
We take $\sigma^2_{\rho_l (r_{li})}$ to be the statistical uncertainty on $\rho_l (r_{l i})$ from shape noise (see Sect.~\ref{sec:method:errors}).
These weights are independent of the shear itself; they depend only on sky position, the radial bin, and lens redshift.
This is a consequence of the fact that the inferred densities $\rho_l$ are linear in the observed shear $G_l$.\footnote{
 This linearity holds with the approximation $G \approx \Delta \Sigma$.
 If instead we take the difference between shear and reduced shear into account, the densities are a non-linear function of shear.
 Then, $\sigma^2_{g_{l i}}$ does depend on the shear, so that the weights $w_{li}$ may be affected by fluctuations in the shear.
 This can be unintuitive to reason about.
 Thus, to determine the weights $w_{li}$ we always use the linear approximation even if we non-linearly convert $G$ to $\Delta \Sigma$.
}
We use the same weights to calculate stacked overdensities $\delta$,
\begin{align}
 \delta_i = \sum_l w_{l i} \frac{\rho_l (r_{li}) - \overline{\rho}_l}{\overline{\rho}_l} \,.
\end{align}
These are not strictly the inverse variance weights for the stacked $\delta$ (which would require an additional factor of $\overline{\rho}_l^2$ in the weights) and therefore lead to slightly increased statistical uncertainties.
We make this trade-off since additional $\overline{\rho}_l^2$ factors in the weights would heavily down-weight low-redshift clusters and hence the clusters where the feature we discuss in Sect.~\ref{sec:results} is most prominent.
In any case, we verified that stacked profiles of $\rho_l (r) - \overline{\rho}_l$, for which the weights we are using \emph{are} the inverse variance weights, have a very similar shape as stacked $\delta_l$ profiles (including the feature discussed below).

To plot the stacked density profiles as a function of radius $r$, we calculate the weighted average of the radii $r_{li}$ (which are numerically equal to the projected radii $R_{li}$ from Eq.~\eqref{eq:R_li}),
\begin{align}
 r_i = \sum_l w_{l i} r_{l i} \,.
\end{align}
Plotting the stacked density profiles as a function of this ``stacked'' radius makes sense when we stack at fixed physical radius $r$.
However, below we often stack at a fixed re-scaled radius $x$, which we take to be $r$ divided by a (proxy for) cluster size.
Mathematically, this means for the $i$-th bin we have bin edges $R_{l i}^\mathrm{min}$ and $R_{li}^\mathrm{max}$ given by $x_i^\mathrm{min} \cdot r_l$ and $x_i^\mathrm{max} \cdot r_l$ where $r_l$ is the cluster size by which we re-scale.
In such a case, it makes more sense to plot the stacked density profiles as a function of a stacked $x$ rather than a stacked $r$, namely
\begin{align}
 x_i = \sum_l w_{l i} x_{l i} \,,
\end{align}
where $x_{l i}$ is defined analogously to Eq.~\eqref{eq:R_li},
\begin{align}
 x_{l i } = \frac{
  \sum_{s \in {l i}} W_{l s} (1 - f_\mathrm{cl}^{l i s}) \cdot (R_{ls}/r_l)
}{
  \sum_{s \in {l i}} W_{l s} (1 - f_\mathrm{cl}^{l i s})
} \,.
\end{align}

\subsection{Error propagation}
\label{sec:method:errors}

We calculate a statistical covariance matrix by linearly propagating the shape noise from the individual shear profiles $G_l$ to the final stacked (over-)density profiles.
In particular, the uncertainty on an individual cluster's $G_l$ due to shape noise is
\begin{align}
 \sigma^2_{G_l (R_{l i})} = B_{li}^2 \frac{
  \sum_{s \in li} W_{ls}^2 \cdot \Sigma_{\mathrm{crit},ls}^2 \sigma^2_{\epsilon/\mathcal{R},b(s)}
 }{
  \left(\sum_{s \in li} W_{l s} \right)^2
 } \,,
\end{align}
where $\sigma^2_{\epsilon/\mathcal{R},b(s)}$ is the shape noise of an individual source $s$ in tomographic bin $b$.
Following \citet{Grandis2024}, we estimate $\sigma^2_{\epsilon/\mathcal{R},b}$ from the observed ellipticity dispersion,
\begin{align}
 \sigma^2_{\epsilon/\mathcal{R},b} = \frac12 \frac{
  \sum_l \sum_{s \in b} W_{ls}^2 \cdot \mathcal{R}_s^{-2} (\epsilon^2_{t,ls} + \epsilon^2_{\times,ls})
 }{
  \sum_l \sum_{s \in b} W_{ls}^2
 } \,,
\end{align}
where $\epsilon_\times$ is the cross component of the ellipticity, the sum over $l$ runs over the whole eRASS:5 sample, and the sum over $s$ runs over all sources in tomographic bin $b$ within $35\,\mathrm{Mpc}$ projected distance around the lenses.

For each cluster, we linearly propagate these statistical uncertainties on the shear profiles $G_l (R_{li})$ into a covariance matrix $\operatorname{Cov}(\delta_l (r_{li}), \delta_l (r_{l j}))$ of the overdensity profiles of the individual clusters (and similarly for the density profiles).
We calculate the necessary Jacobians using automatic differentiation \citep[see][]{Mistele2024b}.
We finally propagate this covariance matrix for individual clusters into the covariance matrix for the stacked overdensity profiles,
\begin{align}
 \operatorname{Cov}(\delta_i, \delta_j) = \sum_l w_{li} w_{lj} \operatorname{Cov}(\delta_l (r_{li}), \delta_l (r_{l j})) \,,
\end{align}
where we assume that the shear measurements around different clusters $l$ and $l'$ are statistically independent.

The above assumes that shape noise is the only contribution to the statistical uncertainty and that different lenses are independent (i.e., they do not overlap).
Below, we therefore also show the covariance matrix obtained from a jackknife procedure, splitting the sample according to a \Healpix pixelization with `NSIDE=8'.
This allows us to estimate the impact of, for example, uncertainties due to fluctuations in the large-scale structure \citep[e.g.][]{Hoekstra2003,Gruen2015,Wu2019}.
Furthermore, our analytical covariance matrix does not apply when using overlapping radial sliding windows to measure the shear.
Thus, when a covariance estimate is required for sliding windows, we use the jackknife covariance.

\subsection{Subtraction of randoms}
\label{sec:method:randoms}

As mentioned above, we subtract the shear measured around random coordinates from the shear measured around the true lenses to remove potential additive shear biases.
In particular, for a given lens $l$ and radial bin $i$, we subtract the shear measured around random sky positions (see Eq.~\eqref{eq:G_l}),
\begin{align}
 \label{eq:G_random}
 G_\mathrm{random}^{li} = \frac{
  \sum_r \sum_{s \in {r i}} W_{r s} \cdot 
  \Sigma_{\mathrm{crit},rs} \mathcal{R}_s^{-1} \epsilon_{t, rs}
}{
  \sum_r \sum_{s \in {r i}} W_{r s}
} \,.
\end{align}
The randoms $r$ should have a similar distribution in redshift and sky position as the real lenses.
In practice, we begin with the 20 catalogs of random coordinates from the Legacy Survey DR10 \citep{Myers2023}.
We assign each of the 20 catalogs a redshift from a uniform grid between $0.1$ and $1.1$.
At each redshift and for each \Healpix pixel with `NSIDE=8', we pick 5200 coordinates from the corresponding catalog in the corresponding pixel.
The number 5200 is chosen to be 100 times the maximum number of eRASS:5 cosmology clusters at any single pixel and redshift.
We then pre-calculate the stacked random shear according to Eq.~\eqref{eq:G_random} separately for each redshift, pixel, and radial bin.
We use 18 logarithmic bin edges between $0.2\,\mathrm{Mpc}$ and $35\,\mathrm{Mpc}$.
The $G_\mathrm{random}^{li}$ we subtract for a given lens and a given radial bin is then obtained from the pre-calculated stacked shears in that lens' \Healpix pixel by linearly interpolating in redshift and logarithmic radius.

The choice of `NSIDE=8' is a compromise between making the pixels large enough to avoid subtracting the signal from the real clusters and making the pixels small enough to avoid washing out information about the precise lens selection function. 
We have experimented with larger \Healpix resolutions of 16, 32, and 64, and found that a progressively larger, positive shear signal is measured from the randoms.
This positive signal grows with radius and likely originates from the (heavily miscentered) shear of the true clusters.
Indeed, at `NSIDE=16', the radius of a pixel is roughly $1.8^\circ$, which corresponds to about $22\,\mathrm{Mpc}$ at $z = 0.2$, and is smaller than the radial range of our shear profiles (which we measure out to $35\,\mathrm{Mpc}$ for the richest clusters).
The `NSIDE=8' randoms' shear is negligible in practice.
This holds for both DES~Y3 and DECADE.
A small effect from subtracting randoms is also in line with previous studies using the DES~Y3 weak lensing data \citep[e.g.,][]{Prat2022}.

The above procedure subtracts the randoms' shear at the level of individual lenses using, for each lens, the randoms in the same region of the sky as that lens.
Due to this and due to the linearity of our deprojection and stacking procedures (Sect.~\ref{sec:method:shear_to_rho} and Sect.~\ref{sec:method:stacking}), we expect that additive shear biases can be successfully subtracted even when they vary with sky position, as long as they vary on scales larger than our \Healpix resolution.
In Sect.~\ref{sec:results:checks:randoms}, we additionally consider an alternative procedure of subtracting randoms, again finding only a small effect.

\section{Results}
\label{sec:results}

Before stacking clusters, we rescale cluster-centric distances by a characteristic measure of cluster size. This is motivated by the expectation in $\Lambda$CDM that density profiles of clusters with different masses become self-similar when expressed in units of their characteristic size \citep{Voit2005}. As an observationally motivated proxy for cluster size, we use $(\lambda/25)^p\,\mathrm{Mpc}$, with $p$ close to $1/3$. This choice is motivated by the scaling of a cluster's radial extent as the cube root of its mass together with the roughly linear scaling between richness $\lambda$ and mass \citep[with an exponent of 1.14, 0.99, and 1.01, respectively, in][]{Saro2015, Capasso2019, Kluge2024, Ghirardini2024}. We therefore stack clusters in terms of the rescaled radius
\begin{align}
 x = \frac{r}{(\lambda/25)^p\,\mathrm{Mpc}} \,,
\end{align}
rather than the physical radius $r$.
This rescaling is particularly crucial when searching for features in the density profiles. Characteristic radii, such as the splashback radius, are expected to scale with cluster size and therefore occur at different physical radii for clusters of different richness. Stacking at a fixed physical radius $r$ may smear out such features (see Appendix~\ref{sec:appendix:more_robustness} below). By rescaling the radius to a radius that scales with cluster size, such features may become aligned in $x$, thereby enhancing their detection in the stacked density profiles \citep[e.g.][]{Bulbul2014, Zhang2025, Bulbul2026}.

\begin{figure}
  \includegraphics[width=\columnwidth]{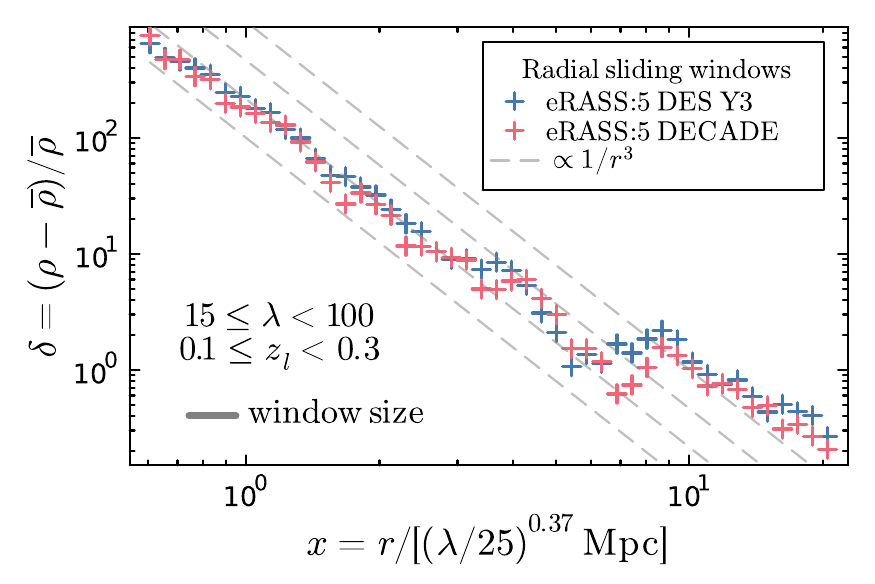}
  \caption{%
  Overdensity profiles of the eRASS:5 clusters in the DES~Y3 (blue) and DECADE (red) footprints.
   The overdensity profiles are stacked as a function of the rescaled radius $x = r/[(\lambda/25)^{0.37}\,\mathrm{Mpc}]$, where $(\lambda/25)^{0.37}\,\mathrm{Mpc}$ is a proxy for cluster size.
   We show results using overlapping, sliding radial windows.
   At $x \sim 6$ we see a non-monotonic feature.
   Apart from this feature, the profiles are qualitatively consistent with the expectations in $\Lambda$CDM: An $r^{-3}$ decline due to the one-halo term followed by a shallower decline corresponding to the two-halo term.
  }
  \label{fig:rho_of_x_windows}
\end{figure}

\begin{figure}
  \includegraphics[width=\columnwidth]{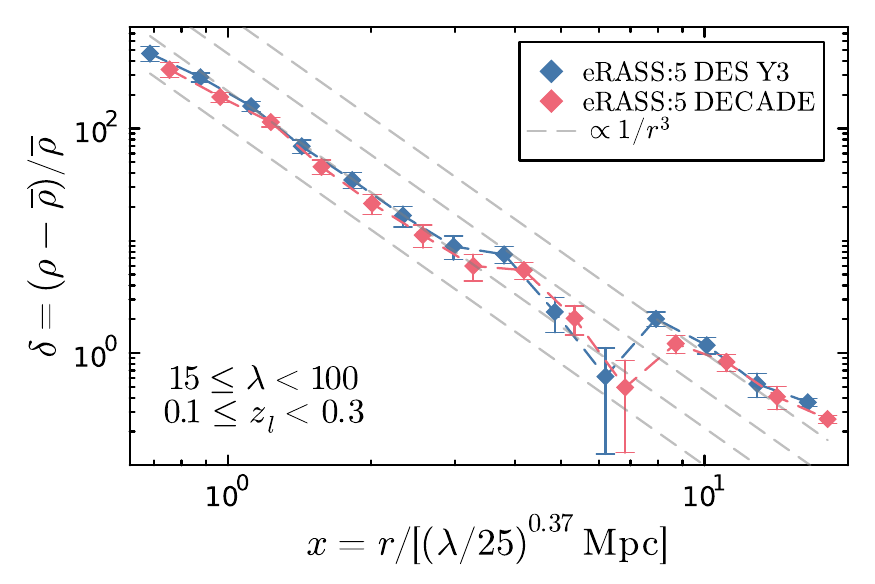}
  \caption{Overdensity profiles as in Fig.~\ref{fig:rho_of_x_windows}, but using non-overlapping radial bins instead of overlapping sliding windows.
  The radial bins are chosen such that one data point is roughly at the middle of the radial range of the feature around $x \sim 6$ as seen using the overlapping radial windows.
  The error bars indicate the uncertainties from shape noise.
  The agreement between the two independent profiles is striking.}
  \label{fig:rho_of_x}
\end{figure}

\begin{figure}
  \includegraphics[width=\columnwidth]{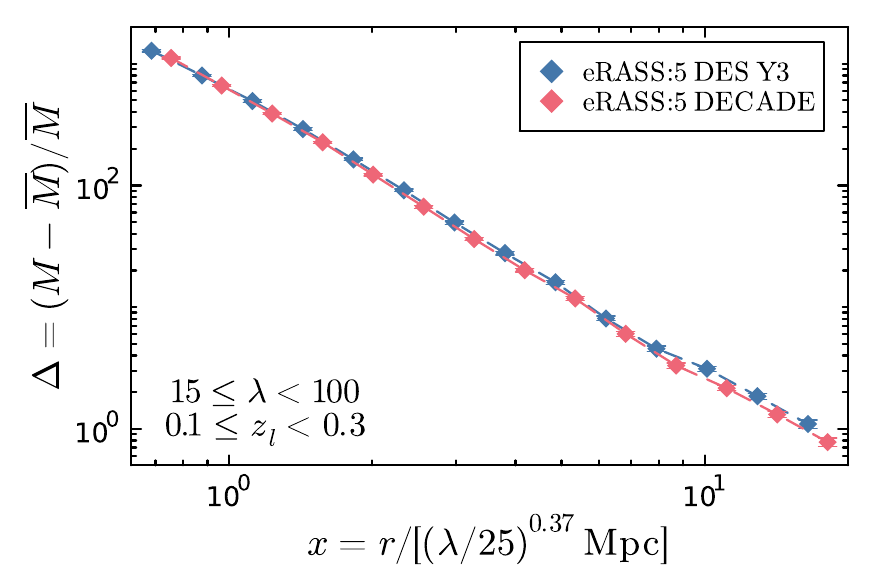}
  \caption{%
    Stacked cumulative overdensity profiles $\Delta = (M - \overline{M})/\overline{M}$, with $\overline{M} = (4\pi/3) r^3 \overline{\rho}_m$, calculated in the same way as the local overdensity profiles $\delta$ in Fig.~\ref{fig:rho_of_x}, except that we use the mass deprojection formula Eq.~\eqref{eq:M_from_ESD} instead of the density deprojection formula Eq.~\eqref{eq:rho_from_ESD}.%
  }
  \label{fig:Delta_of_x}
\end{figure}

\begin{figure}
  \includegraphics[width=\columnwidth]{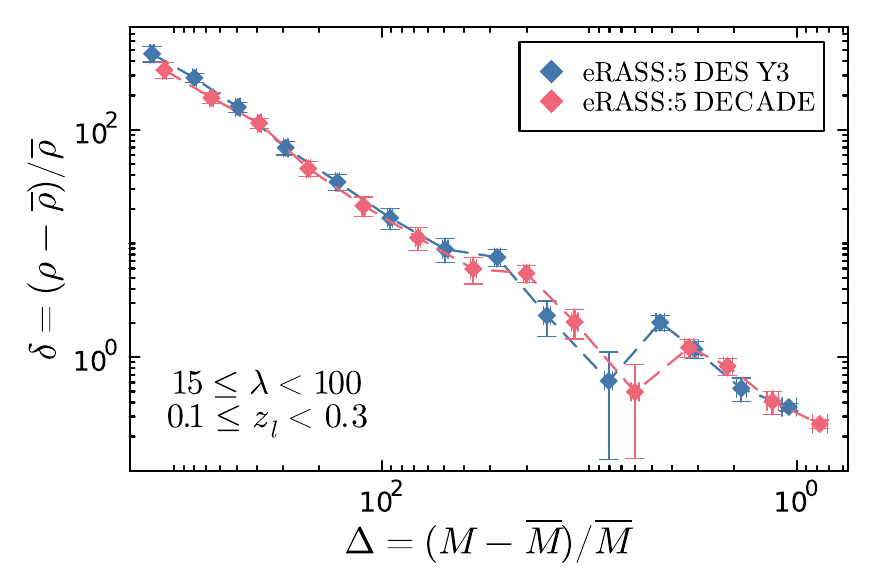}
  \caption{%
    The (local) overdensity profiles $\delta(x) = (\rho(x) - \overline{\rho})/\overline{\rho}$ from Fig.~\ref{fig:rho_of_x} plotted against the (cumulative) overdensity profiles $\Delta(x) = (M(x) - \overline{M}(x))/\overline{M}(x)$ from Fig.~\ref{fig:Delta_of_x}.
    At the minimum of the $x \sim 6$ feature, we have $\Delta \approx 8.1$ for DES and $\Delta \approx 6.0$ for DECADE.
  }
  \label{fig:delta_vs_Delta}
\end{figure}

Our goal in this work is to identify features in the outskirts of clusters; therefore, the choice of radial binning becomes important. While wider radial bins reduce statistical noise, they can also smooth out localized features. To mitigate this, we first perform the analysis using overlapping sliding windows (see Sect.~\ref{sec:method:shear}) and use these results to optimize the choice of radial bins for the final measurements. The overlapping windows produce highly correlated data points and should therefore be interpreted with care, particularly when inspecting the profiles visually. Regardless, they provide a useful detection tool by reducing the likelihood of missing localized features.

\begin{figure*}
  \centering
  \includegraphics[width=1.8\columnwidth]{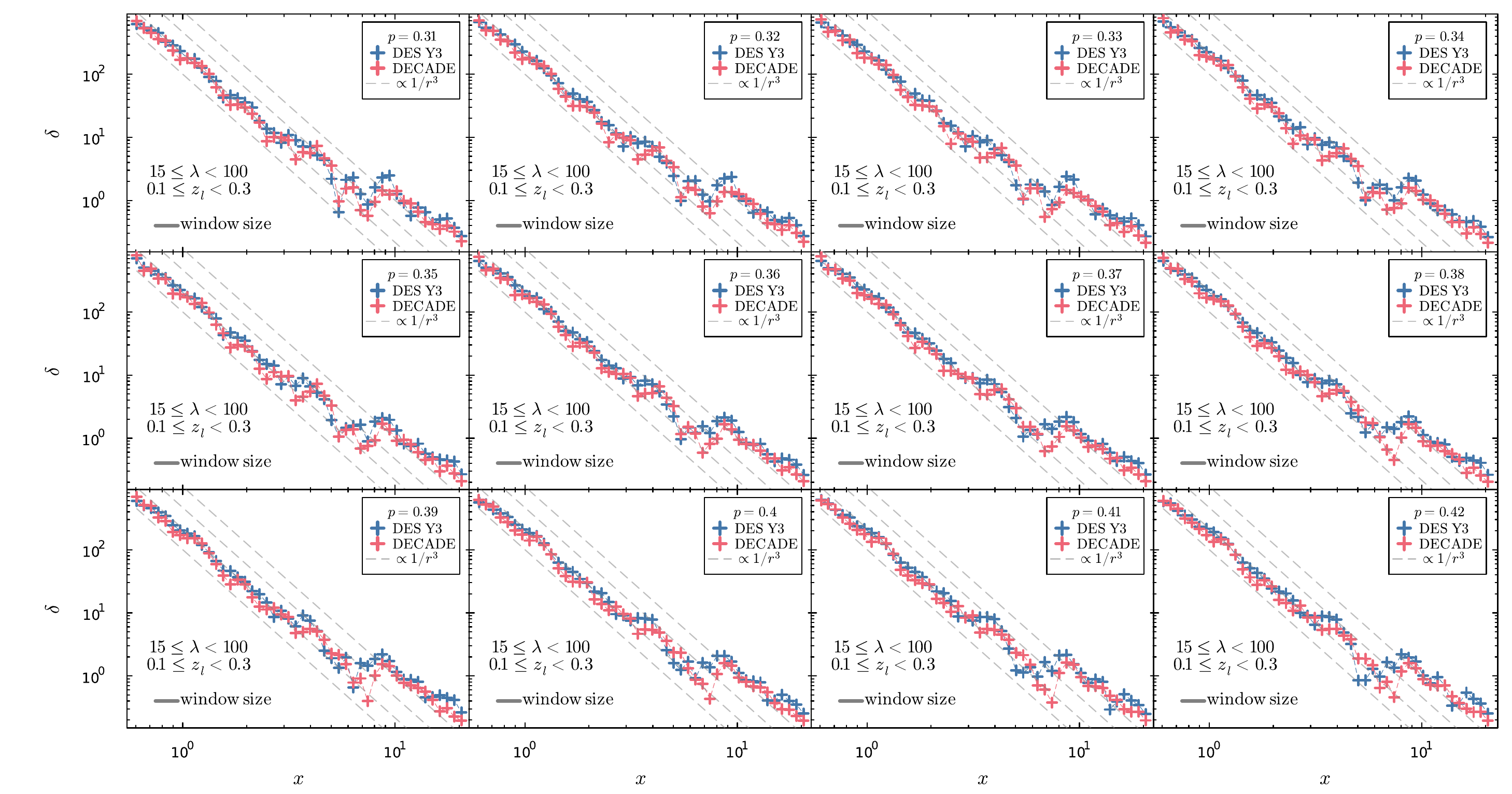}
  \caption{%
   Overdensity profiles using overlapping sliding windows as in Fig.~\ref{fig:rho_of_x_windows} for various powers $p$ in the re-scaled radius $x = r/[(\lambda/25)^p\,\mathrm{Mpc}]$.
   For $p \lesssim 0.36$, we see a double-dip around $x \sim 6$.
   These combine into a single one at $p \geq 0.37$, which is why we choose $p = 0.37$ for most analyses.
  }
  \label{fig:lambda_power_grid}
\end{figure*}

Figure~\ref{fig:rho_of_x_windows} shows the stacked overdensity profiles of the eRASS:5 clusters as a function of the rescaled radius $x$, adopting $p=0.37$.
Given the substantial scatter in the relation between mass (and hence radius) and richness, we explored the range $p=0.31-0.42$, finding that the feature we discuss below becomes most consistent and pronounced for $p=0.37$.
The feature persists over a range of $p$ values around $1/3$ and is therefore not specific to this choice (see Fig.~\ref{fig:lambda_power_grid}).
We separately show results for clusters within the DES and DECADE footprints. Since the two footprints do not overlap, the corresponding cluster samples and the lensing data are statistically independent to leading order. We initially focus on the low-redshift subsample, $z_l < 0.3$, and discuss the higher-redshift clusters separately below.
For later analyses, we choose radial bins such that one data point lies roughly in the middle of the feature's radial range, as observed with the sliding windows.
In particular, for DECADE, the feature location appears slightly shifted relative to DES, so our radial bin edges for DECADE are $10\%$ larger than for DES.
Figure~\ref{fig:rho_of_x} shows results using these radial bins.
They are consistent with the results using radial windows and they are easier to interpret since the data points are less correlated.

At $x \lesssim 3$, the profiles show an approximate $r^{-3}$ decline, broadly consistent with the behavior of the one-halo term in $\Lambda$CDM. At larger radii, around $\delta \sim 5$--$10$, the profiles transition to a shallower slope consistent with the onset of the two-halo regime.
Before this transition, $\Lambda$CDM predicts the splashback radius, the location of the steepest slope.
We do not see a strong steepening in our results in this region, but our results are consistent with a splashback radius as expected from $\Lambda$CDM, see the further discussion in Sect.~\ref{sec:discussion}.
At even larger radii, in the infall region, we identify a distinct feature around $x \sim 6$ that is not expected from the standard $\Lambda$CDM profile.
Apart from a small shift in its location, the feature is consistent between DES and DECADE, providing a first indication that it is not merely a statistical fluctuation.
The location of the feature matches that of the turnaround radius in $\Lambda$CDM \citep{Pavlidou2014,Korkidis2025}.
Indeed, the turnaround radius sits at a mass-independent location in terms of the ``cumulative'' overdensity $\Delta(r) = (M(r) - \overline{M}(r))/\overline{M}(r)$ with $\overline{M}(r) = (4\pi/3) \, r^3 \overline{\rho}_m$ (as opposed to the local overdensity $\delta$ we have discussed so far).
The stacked $\Delta$ is shown in Fig.~\ref{fig:Delta_of_x}.
At the minimum of the feature, we find $\Delta \approx 8.1$ for DES and $\Delta \approx 6.0$ for DECADE.
These are slightly lower than the expected values at the turnaround radius, which range from about $\Delta \approx 9.5$ to about $\Delta \approx 7.7$ for redshifts between $0.1$ and $0.3$ and assuming a flat $\Lambda$CDM cosmology with $\Omega_m = 0.3$ \citep{Tanoglidis2016,Ghirardini2024,Planck2018}.
However, $\Delta(r)$ drops rapidly with radius, and these theoretical values are well within the radial range covered by the feature (see Fig.~\ref{fig:delta_vs_Delta}).
We further discuss a possible relation to the turnaround radius in Sect.~\ref{sec:discussion} below.

\begin{figure}
  \includegraphics[width=\columnwidth]{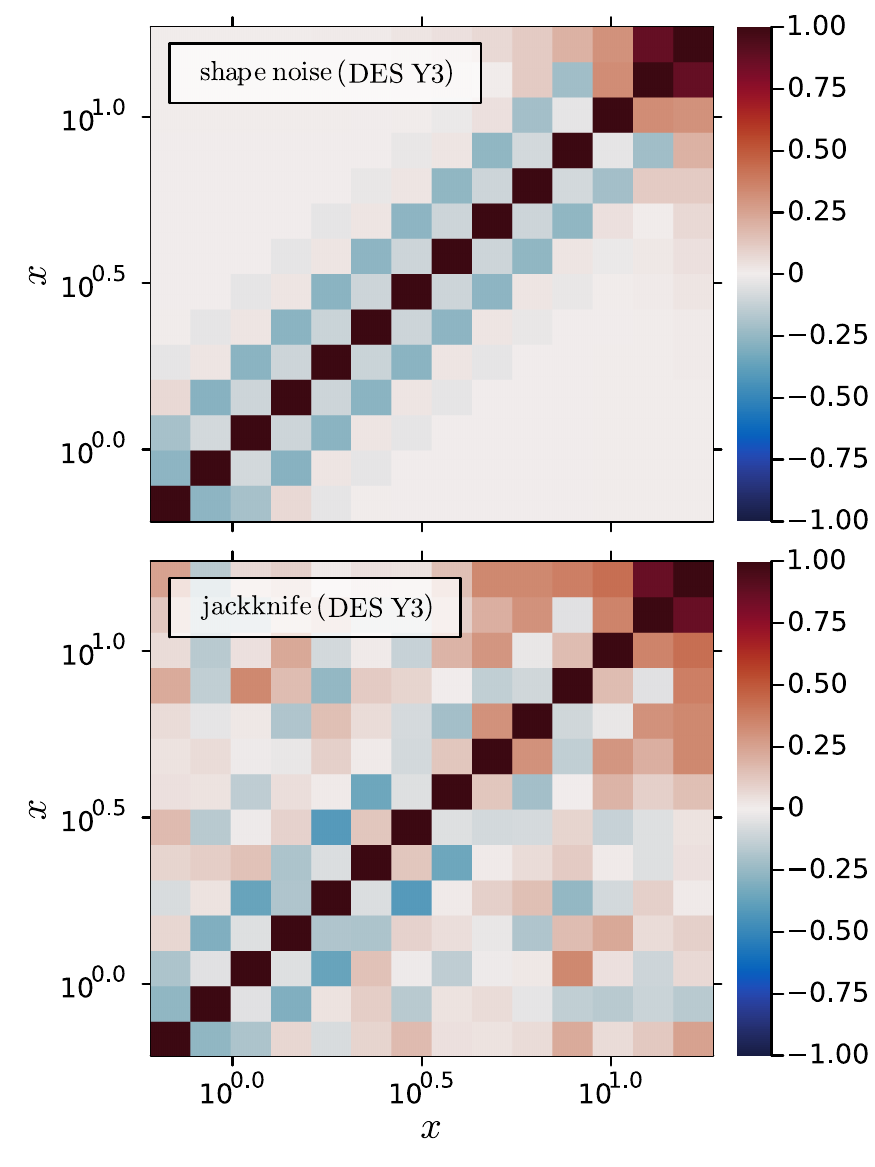}
  \caption{%
   Correlation matrices for the DES~Y3 overdensity profile from Fig.~\ref{fig:rho_of_x}.
   The correlation matrices for DECADE are similar.
   \emph{Top}: Correlation matrix calculated from only the shape noise. We see a clear structure with neighboring data points being (anti-)correlated.
   The change of this pattern at the largest radii is because the observed shear profiles must be extrapolated beyond the last data point when converting them to density profiles (Sect.~\ref{sec:method:shear_to_rho}).
   \emph{Bottom}: Jackknife correlation matrix calculated from an `NSIDE=8` \Healpix pixelization. The level of (anti-)correlation remains similar, but the clear pattern we see in the top panel is partially washed out.
  }
  \label{fig:shapevsjack_correlation}
\end{figure}

\begin{figure}
  \includegraphics[width=\columnwidth]{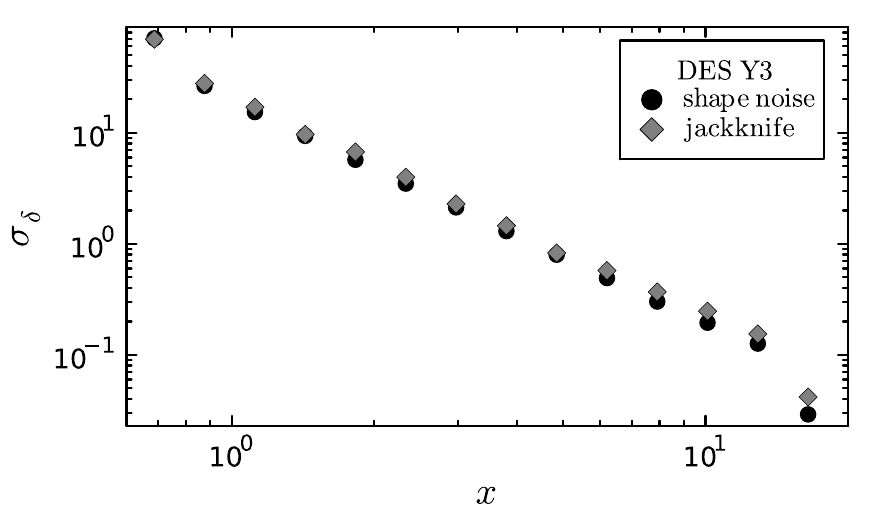}
  \caption{%
   Uncertainties of the overdensity profile for DES~Y3 from Fig.~\ref{fig:rho_of_x} as estimated using shape noise and using a jackknife covariance matrix.
   At large radii, the jackknife uncertainties exceed the shape noise uncertainties.
   Results for DECADE are similar.
  }
  \label{fig:shapevsjack_sigma}
\end{figure}

The error bars shown above represent the statistical uncertainties due to shape noise.
Neighboring data points are not entirely independent, not even in the binned analysis, due to the integrals in Eq.~\eqref{eq:rho_from_ESD} which mix shear measurements from different radii.
This can be seen from the off-diagonal components in the DES correlation matrix shown in the top panel of Fig.~\ref{fig:shapevsjack_correlation}.
The DECADE correlation matrix is similar.
The (anti-)correlation typically remains below $0.3$ and spans two neighboring radial bins.
For the outermost few radial bins, the correlation becomes stronger, which is because the extrapolation of the shear profiles becomes important there (Sect.~\ref{sec:method:shear_to_rho}).
In practice, there may be contributions to the covariance matrix other than shape noise, for example, fluctuations due to the large-scale structure \citep[e.g.][]{Hoekstra2003,Gruen2015,Wu2019}.
To estimate their effect, the lower panel of Fig.~\ref{fig:shapevsjack_correlation} shows a jackknife correlation matrix obtained using an `NSIDE=8' \Healpix pixelization.
Figure~\ref{fig:shapevsjack_sigma} shows the effect on the uncertainties, which is non-zero but moderate.
The jackknife procedure washes out some of the clear structure in the correlation matrix.
This may be due to both the jackknife correlation being somewhat noisy and the negative correlations being canceled by the positive correlations induced by fluctuations in the large-scale structure.

\begin{figure}
  \includegraphics[width=\columnwidth]{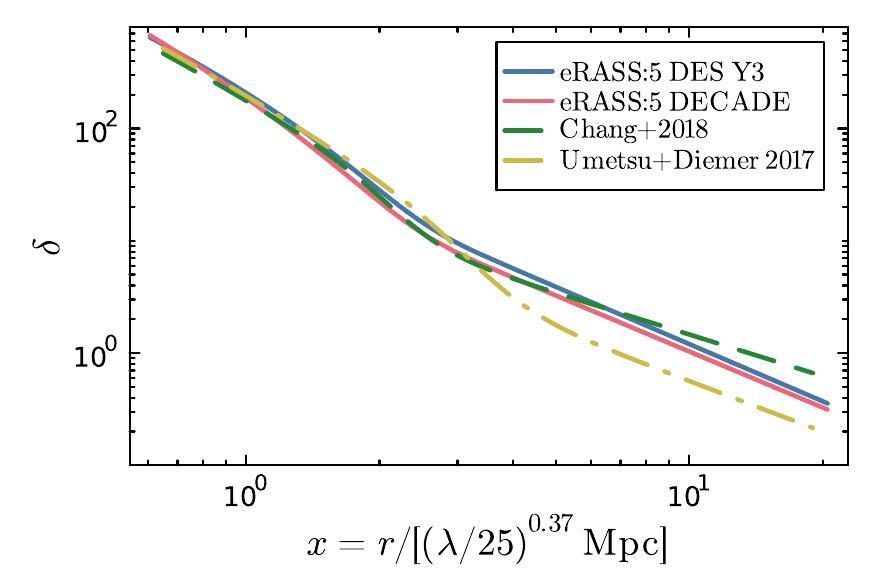}
  \caption{%
   Best-fit overdensity profiles following the parametric form from \citet{Diemer2023} for DES Y3 (blue) and DECADE (red).
   For reference, we also show fits using the \citet{Diemer2014} form from \citet{Umetsu2017} (analyzing the CLASH clusters) and \citet{Chang2018} (analyzing the DES Y1 clusters); their profiles are rescaled to roughly match the normalization of our results in the one-halo region, enabling a direct comparison of their radial shape.
  }
  \label{fig:literature}
\end{figure}

Previous lensing analyses did not detect a feature like the one we find at $x \sim 6$ \citep[e.g.,][]{Umetsu2017,Chang2018,Contigiani2019,Mistele2025}.
This may be expected, since these analyses cover a more limited radial range, use smaller cluster samples, and, in most cases, are based on fitting parametric density profiles whose forms are not designed to detect non-monotonic features.
As examples, Figure~\ref{fig:literature} shows the fit results from \citet{Umetsu2017} and \citet{Chang2018} next to parametric fits to our overdensity profiles from Fig.~\ref{fig:rho_of_x_windows}.
\citet{Umetsu2017} analyze the CLASH clusters while \citet{Chang2018} study the DES Y1 clusters.
We use the parametric form from \citet{Diemer2023}, whereas \citet{Umetsu2017} and \citet{Chang2018} use the older \citet{Diemer2014} form.
We remove the region around the $x \sim 6$ feature from our fits since the feature clearly does not follow that parametric form.
We further remove the outermost three data points since they may be affected by extrapolation of the shear profiles (see Sect.~\ref{sec:results:checks}).
We rescale the fits from the literature to roughly match the normalization of our results in the one-halo region, thus enabling a direct comparison of the radial shape of the profiles.
All fits roughly agree on the slope in the one-halo region.
The fits to our eRASS:5 results roughly reproduce the location of the splashback radius, defined as the steepest slope.
Namely, we find the steepest slope in our best fits at $x = 1.9$ for DES Y3 and at $x = 1.8$ for DECADE.
From Fig.~\ref{fig:Delta_of_x} we see that this is within $15\%$ of $r_{200m}$, which is compatible with theoretical expectations \citep{Diemer2014}.
Differences exist in the precise shape of the transition to the two-halo term and in the normalization and slope of the two-halo term.
For example, the CLASH result lies well below our results in the two-halo region.
This may be explained by the fact that the CLASH data only weakly constrain the two-halo regime.
Indeed, a larger outer amplitude, close to our results, is allowed by the posterior uncertainties of \citet{Umetsu2017}.
Thus, there is likely no real disagreement over the normalization of the two-halo term.
Similarly, our posteriors for the slope of the two-halo term overlap with those of \citet{Umetsu2017} and \citet{Chang2018}.
In contrast, the feature we find at $x \sim 6$ is a significant novelty in our non-parametric results, illustrating the power of our deprojection technique in detecting non-monotonic features in stacked density profiles.

\begin{figure}
  \includegraphics[width=\columnwidth]{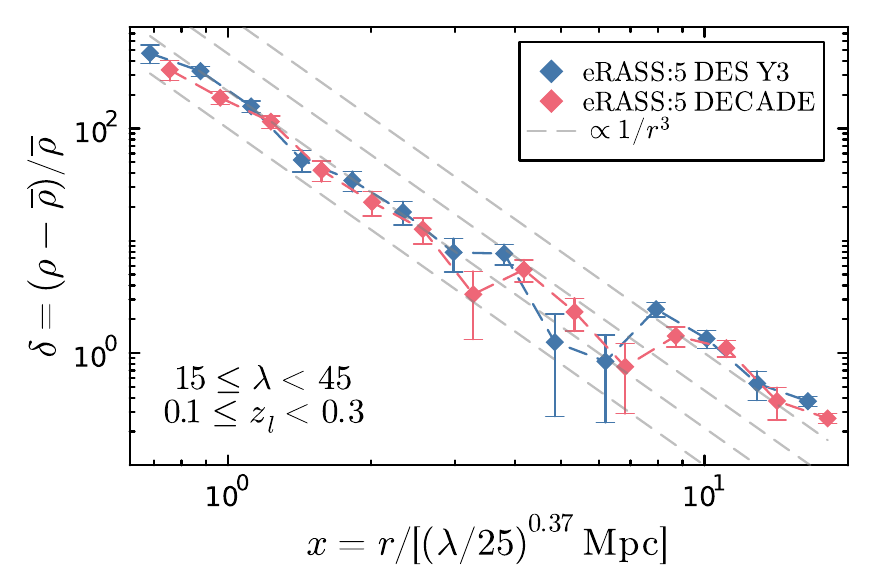}
  \caption{%
   Same as Fig.~\ref{fig:rho_of_x}, but including only low-richness clusters.
   The feature at $x \sim 6$ is still clearly visible.
   }
  \label{fig:rho_of_x_lowlambda}
\end{figure}

\begin{figure}
  \includegraphics[width=\columnwidth]{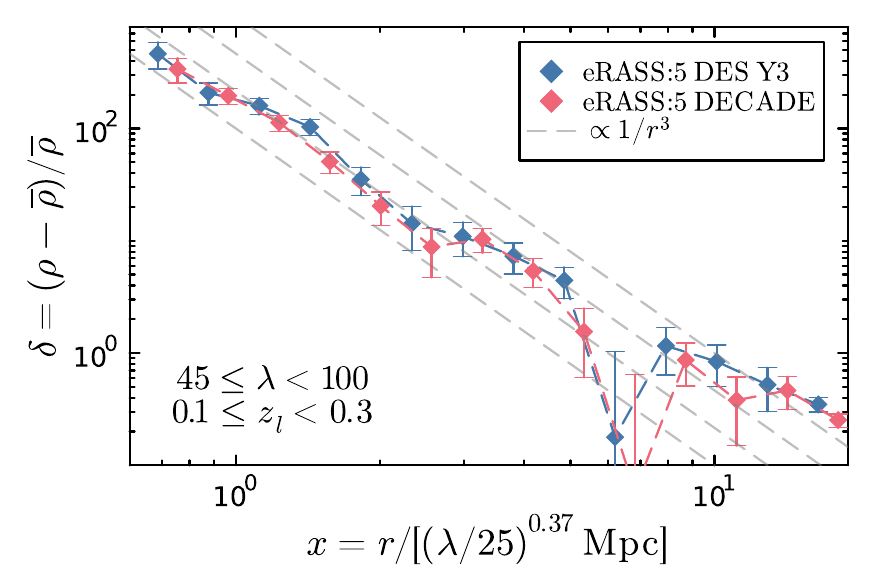}
  \caption{%
   Same as Fig.~\ref{fig:rho_of_x}, but including only high-richness clusters.
   The feature at $x \sim 6$ is still clearly visible.%
   }
  \label{fig:rho_of_x_highlambda}
\end{figure}

\subsection{Mass dependence}

The location of the turnaround radius in $\Lambda$CDM scales as $M^{1/3}$ where $M$ is the mass enclosed within the turnaround radius \citep{Pavlidou2014}.
The $x \sim 6$ feature is consistent with this scaling, since it becomes visible after rescaling cluster-centric distances by a proxy that scales similarly to $M^{1/3}$:
Indeed, as mentioned above, $\lambda$ scales roughly linearly with mass, and our proxy for cluster size is proportional to $\lambda^p$ with $p$ close to $1/3$.
Conversely, in Appendix~\ref{sec:appendix:more_robustness}, we show that the feature becomes smeared out when we omit this rescaling, which is again consistent with an $M^{1/3}$ scaling of the feature's location.

To further test this scaling, we split our sample into high- and low-richness bins.
If the feature's location indeed scales with $M^{1/3}$, and if $(\lambda/25)^p$ is a good proxy for $M^{1/3}$, the feature should remain at the same rescaled radius $x$ in both subsamples.
Figure~\ref{fig:rho_of_x_lowlambda} and Fig.~\ref{fig:rho_of_x_highlambda} show the results for the low and high richness bins, respectively.
The feature persists despite the increased noise level from the smaller sample size and remains roughly consistent between DES and DECADE.
In addition, there are hints of a secondary feature at $x \sim 2-3$ which may, however, just be statistical fluctuations.
In DES, the location of the $x \sim 6$ feature appears to shift slightly toward lower values of $x \sim 5-6$ in the low-richness sample.
This may be due to statistical fluctuations or it may be a hint that the feature doesn't perfectly match the expected $M^{1/3}$ scaling of the turnaround radius.

\begin{figure}
  \includegraphics[width=\columnwidth]{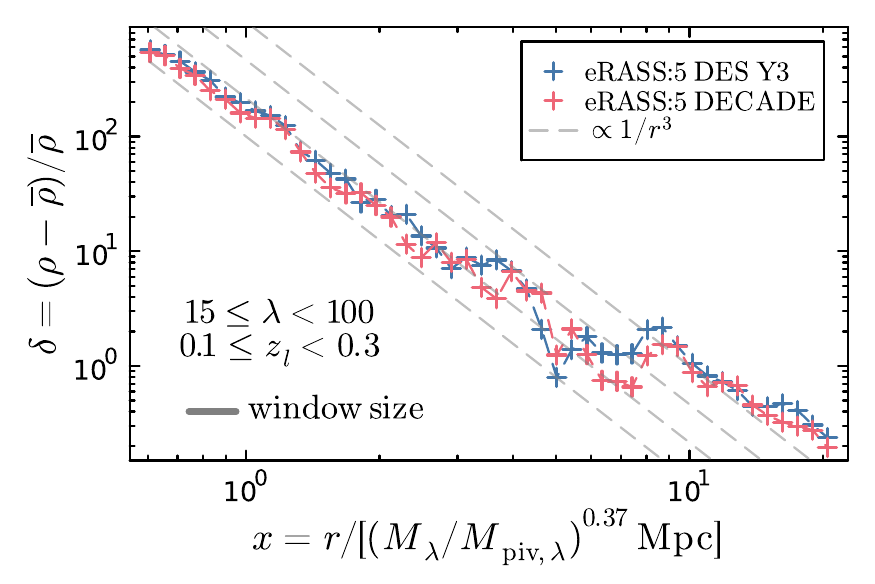}
  \caption{%
   Overdensity profiles using sliding windows as in Fig.~\ref{fig:rho_of_x_windows}, but the radii $r$ are re-scaled with $(M_\lambda/M_{\mathrm{piv},\lambda})^{0.37}$ instead of $(\lambda/25)^{0.37}$.
    Here, $M_\lambda$ is the mass implied by the mass-richness scaling relation from \citet{Ghirardini2024,Kluge2024} and $M_{\mathrm{piv},\lambda} = 2 \cdot 10^{14}\,M_\odot$.
    This alternative cluster-size proxy yields results similar to our fiducial one, including a clear feature around $x \sim 6$.
   }
  \label{fig:rho_x_MrescaledKluge24}
\end{figure}

\begin{figure}
  \includegraphics[width=\columnwidth]{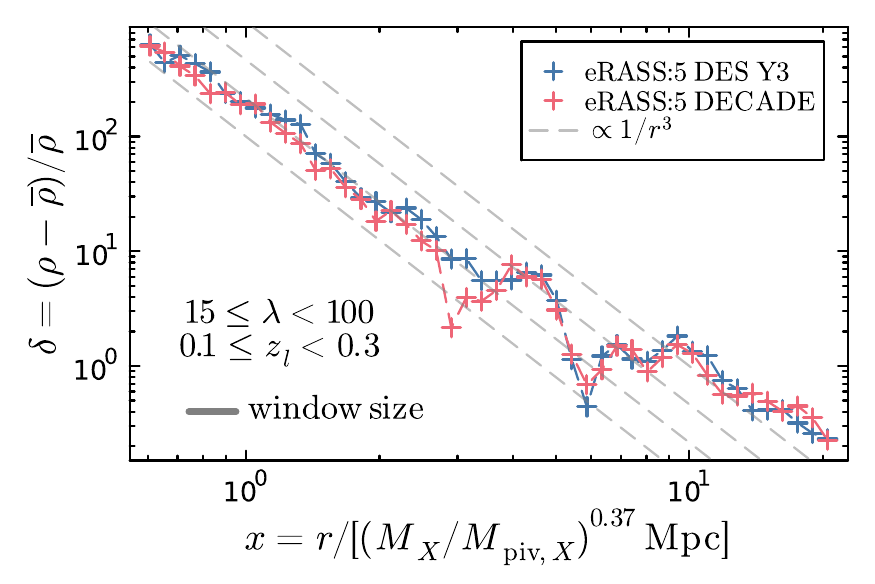}
  \caption{%
   Overdensity profiles using sliding windows as in Fig.~\ref{fig:rho_of_x_windows}, but the radii $r$ are re-scaled with $(M_X/M_{\mathrm{piv},X})^{0.37}$ instead of $(\lambda/25)^{0.37}$.
    Here, $M_X$ is the mass implied by the mass-$L_X$ scaling relation from \citet{Bulbul2019} and $M_{\mathrm{piv},X} = 1.5 \cdot 10^{14}\,M_\odot$.
    This alternative cluster-size proxy yields results similar to our fiducial one, including a clear feature around $x \sim 6$.
    In DECADE, there is a hint of a secondary feature at $x \sim 3$, similar to Fig.~\ref{fig:rho_of_x_lowlambda}.
   }
  \label{fig:rho_x_MrescaledBulbul19}
\end{figure}

As another test of whether the location of the $x \sim 6$ feature scales with $M^{1/3}$, we employ two different observable-mass scaling relations from the literature.
We first replace our proxy for cluster size with a proxy based on the mass $M_\lambda$ implied by the eRASS1 richness-mass scaling relation \citep{Ghirardini2024,Kluge2024}.
In particular we adopt $(M_\lambda/M_{\mathrm{piv},\lambda})^{0.37}\,\mathrm{Mpc}$ with $M_{\mathrm{piv},\lambda} = 2 \cdot 10^{14}\,M_\odot$ as our proxy.
The results are shown in Fig.~\ref{fig:rho_x_MrescaledKluge24}.
We next employ a proxy based on the mass $M_X$ implied by the mass-$L_X$ relation from \citet{Bulbul2019}.
This mass-$L_X$ relation is calibrated using SZ-selected clusters from the South Pole Telescope \citep{Bleem2015}, and we expect it to be approximately valid for our sample since it covers a similar range in mass and redshift.
In particular, we adopt $(M_X/M_{\mathrm{piv},X})^{0.37}\,\mathrm{Mpc}$ with $M_{\mathrm{piv},X} = 1.5 \cdot 10^{14}\,M_\odot$ as our proxy.
The results are shown in Fig.~\ref{fig:rho_x_MrescaledBulbul19}.
The values of $M_{\mathrm{piv},\lambda}$ and $M_{\mathrm{piv},X}$ are chosen such that all three of our proxies roughly agree on the implied cluster size.
We find that, for DECADE, the $M_X$-based proxy shows a hint of a secondary feature at $x \sim 3$, similar to what we saw in Fig.~\ref{fig:rho_of_x_lowlambda} in the low-richness sample.
Apart from that, all three proxies clearly show a feature around $x \sim 6$, consistent between DES and DECADE.
This is consistent with the feature's location matching the $M^{1/3}$ scaling of the turnaround radius.
The precise shape of the feature varies with the proxy for cluster size used.
One possible reason is that the feature has a somewhat different shape around each cluster, reflecting the complexity of the infall region, so that slightly different rescalings with varying scatter may cause features from individual clusters to overlap differently in the stack.

\begin{figure}
  \includegraphics[width=\columnwidth]{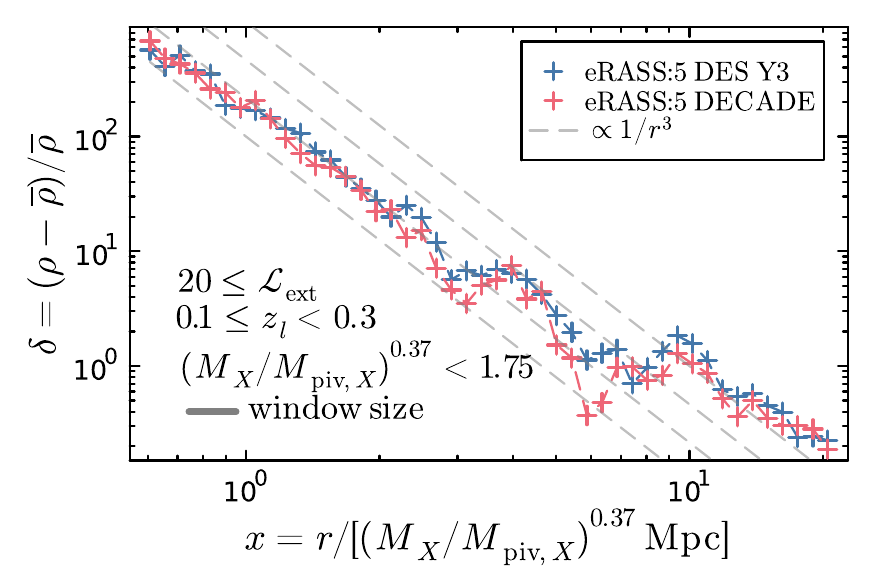}
  \caption{%
   Overdensity profiles using sliding windows as in Fig.~\ref{fig:rho_of_x_windows}, but using as little optical information as possible.
   Specifically, we rescale the x-axis by the $M_X$ mass proxy (Fig.~\ref{fig:rho_x_MrescaledBulbul19}), we replace the $\lambda \geq 15$ cut with a cut on extent likelihood, $\mathcal{L}_\mathrm{ext} > 20$, and we replace the $\lambda < 100$ cut with $(M_X/M_{\mathrm{piv},X})^{0.37} < 1.75$.
  }
  \label{fig:rho_of_x_nolambda}
\end{figure}

\begin{figure}
  \includegraphics[width=\columnwidth]{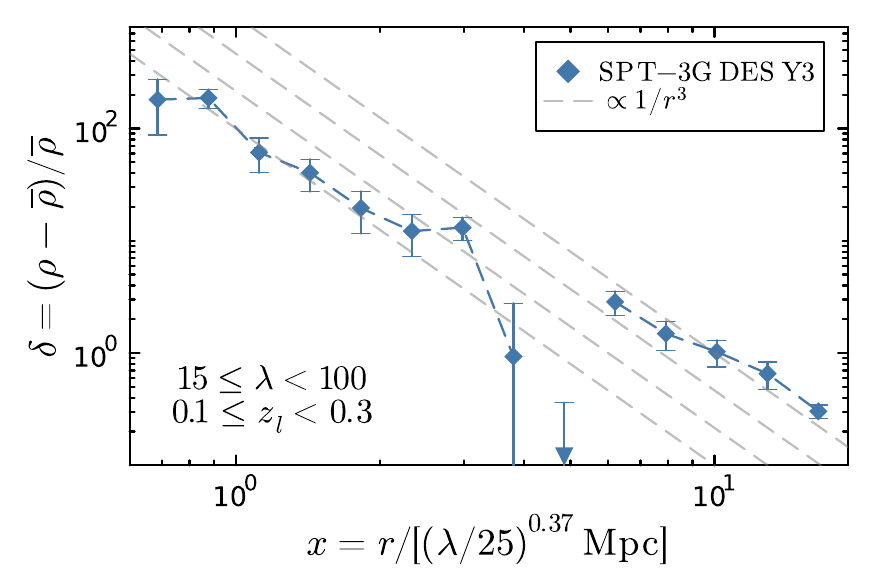}
  \caption{%
   Overdensity profiles for the SPT-3G clusters in DES~Y3.
   Triangles indicate negative data points.
   For simplicity, we do not subtract randoms or correct for contamination from cluster members.
   Uncertainties are larger than with eRASS:5 in the same redshift and richness range.
   At $x \sim 5$, we see a feature similar to the $x \sim 6$ one in eRASS:5.
  }
  \label{fig:rho_of_x_SPT}
\end{figure}

\subsection{Selection effects}
\label{sec:results:selection}

The eRASS:5 clusters are X-ray selected.
For our results above, we additionally applied a richness cut and we re-scaled radii using a cluster size proxy based on richness.
To check for optical selection and projection effects due to this procedure, we here remove these uses of optical information in favor of X-ray-based quantities.
In particular, we use the $M_X$-based cluster size proxy from Fig.~\ref{fig:rho_x_MrescaledBulbul19},  we replace the $\lambda \geq 15$ cut with an extent likelihood cut, $\mathcal{L}_\mathrm{ext} \geq 20$, and we replace the $\lambda < 100$ cut with $(M_X/M_{\mathrm{piv},X})^{0.37} < 1.75$.
As shown in Fig.~\ref{fig:rho_of_x_nolambda}, we find similar results as when using richness cuts and the $M_X$-based size proxy, except that the secondary feature around $x \sim 3$ is now visible in both DES and DECADE.
This shows that the feature at $x \sim 6$ is not artificially induced by optical selection.

To further check for selection effects, we consider the SZ-selected cluster catalog from SPT-3G \citep{Bleem2026}.
For simplicity, we do not subtract randoms, and we do not correct for cluster member contamination (both of which are unimportant to our results as discussed in Sect.~\ref{sec:results:checks:randoms} and Appendix~\ref{sec:appendix:more_robustness}).
Since SPT-3G mostly overlaps with DES rather than DECADE, we consider only DES here.
In particular, we obtain shear from DES Y3 for 463 SPT-3G clusters in the same redshift and richness range used in our main eRASS:5 analysis.
Cross-matching based on sky position within $1.5'$ implies that 293 of these clusters are also in our eRASS:5 sample.
The two samples are therefore not statistically independent but follow distinct selection functions.
The results are shown in Fig.~\ref{fig:rho_of_x_SPT}.
Due to the smaller number of clusters, the uncertainties are larger than in eRASS:5.
Despite this, we find results consistent with a $r^{-3}$ decline in the one-halo region and a non-monotonic feature in the two-halo region.
The minimum of the feature in SPT-3G occurs at $x \sim 5$, which is somewhat lower than the $x \sim 6$ found in eRASS:5.
This may be explained by the fact that the SPT-3G richnesses are typically about a factor of $2$ higher in the clusters that are shared between SPT-3G and eRASS:5, which moves $x = 6$ to about $x = 6/2^{0.37} \approx 4.6$, consistent with our SPT-3G results.
Thus, we find no evidence of selection effects in this analysis of an SZ-selected sample.

\begin{figure}
  \includegraphics[width=\columnwidth]{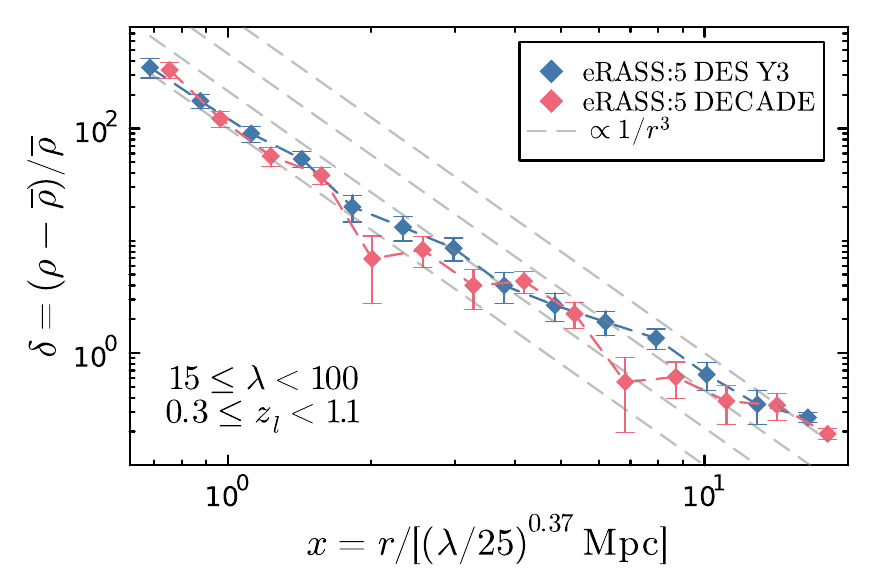}
  \caption{%
   Same as Fig.~\ref{fig:rho_of_x}, but for higher redshifts, $0.3 \leq z_l < 1.1$.
   We do not see a clear and consistent feature.
   This is the reason we generally apply a $z_l < 0.3$ cut in this section.%
   }
  \label{fig:rho_of_x_highz}
\end{figure}

\subsection{Redshift dependence}

Above, we have applied a $z_l < 0.3$ cut.
At higher redshifts, we do not see a clear and consistent feature.
Indeed, Fig.~\ref{fig:rho_of_x_highz} shows results for $0.3 \leq z_l < 1.1$.
We do see a few possible dips, but these are not consistent between DES and DECADE and could be due to the higher noise level at higher redshifts.
It is possible that the location of the feature is redshift dependent and that we should therefore scale $r$ not just by a function of $\lambda$ but also by a function of $z_l$.
We have, however, not found any significant features by making the rescaling redshift dependent.
For example, the cluster size proxies based on the richness-mass and $L_X$-mass scaling relations from \citet{Ghirardini2024,Kluge2024} and \citet{Bulbul2019} (Fig.~\ref{fig:rho_x_MrescaledKluge24} and Fig.~\ref{fig:rho_x_MrescaledBulbul19}), which include a redshift dependence, do not indicate a feature at higher redshifts.
We have similarly not found a feature at high redshift when including a factor of $1+z_l$ to convert physical to comoving radii, or when restricting the redshift range to $z_l < 0.6$.

\subsection{Statistical significance}
\label{sec:results:significance}

To estimate the significance of the observed feature at $x \sim 6$, we fit a broken power law to the data and generate $50,000,000$ mock realizations by adding noise consistent with the observed covariance matrix. The motivation for a broken power law is that, to first order and in $\Lambda$CDM, we expect a $r^{-3}$ decline in the one-halo region followed by a shallower decline in the two-halo region. We then count the fraction of realizations that exhibit a feature at least as strong and as consistent between DES and DECADE as the observed one. The precise procedure is described in Appendix~\ref{sec:appendix:significance}. The resulting frequencies are $14\times10^{-8}$ for a broken power law and $36\times10^{-8}$ for a smoothed broken power law, corresponding to significances of $5.3\sigma$ and $5.1\sigma$, respectively.
Additionally, we have tried excluding the $x \sim 6$ region when fitting the broken power-law profile, since it clearly does not follow a broken power law.
In this case, none of the $50,000,000$ realizations met our criteria, indicating even greater statistical significance.
We have further repeated this procedure with the \citet{Diemer2023} parametric fits from Fig.~\ref{fig:literature}.
We find a significance of $5.2\sigma$ when excluding the $x \sim 6$ region from the fit (as in Fig.~\ref{fig:literature}) and $5.0\sigma$ when not excluding that region.

These results depend to some degree on the choice of fit model and the choice of conditions reflecting the salient features in the observed data.
In addition, they do not take into account all possible look-elsewhere effects, for example, arising from our specific choice of rescaling power $p$ or from the cuts applied to the cluster sample.
Thus, the precise numbers should not be taken to be an absolute and unique assessment of the statistical significance of the feature in our observed data.
Our procedure does, however, highlight that our results are unlikely to be just due to statistical fluctuations.
We also emphasize that the results of the various robustness checks in Sect.~\ref{sec:results:checks} and in Appendix~\ref{sec:appendix:more_robustness} are not included in the significance calculation.
For example, our results for DES remain almost unchanged when restricting the lensing sources to the `blue shear' subsample from \citet{McCullough2024}.
Similarly, our results remain consistent across various rescaling powers $p$.
If taken into account, these additional results are likely to, at least partially, counter potential look-elsewhere and similar effects.

\section{Validation of deprojection method}
\label{sec:method:hydro}

In this section, we validate our deprojection method using the Flamingo suite of hydrodynamic simulations in $\Lambda$CDM \citep{Flamingo2023a,Flamingo2023b,Flamingo2026}.
In particular, we use the `L1\_m9' simulation, generate shear profiles loosely following the procedure of \citet{Grandis2021}, apply our method to these shear profiles, and finally compare the resulting overdensity profiles to the true overdensity profiles.

We calculate azimuthally averaged reduced tangential shear profiles out to $25\,\mathrm{Mpc}$ from projected surface density maps with a projection depth of $\pm 50\,\mathrm{Mpc}$.
Since this projection depth is small compared to the comoving distance to the clusters, we neglect redshift evolution along the line of sight and work in physical distances.
The surface density maps have quadratic cells with a width of about $22.2\,\mathrm{kpc}$.
We consider the particles corresponding to dark matter, gas, and stars.
We neglect the small contributions from black holes and neutrinos.
In this subsection, we adopt $H_0 = 68.1\,\mathrm{km}\,\mathrm{s}^{-1}\,\mathrm{Mpc}^{-1}$ and $\Omega_m = 0.305$, matching the cosmology assumed in the `L1\_m9' simulation.
We use the shear profiles of all 7,887 halos with $M_{200c} \geq 10^{14}\,M_\odot$ from the snapshot at $z=0.2$.
This mass range roughly matches our eRASS:5 sample of massive clusters with $\lambda \geq 15$.
For each cluster, we use three shear profiles obtained by projecting along the three Cartesian axes (the $x$, $y$, and $z$ axes used in the simulations).
Since the difference between shear and reduced shear is not important for our results in this subsection, the choice of source redshift distribution is also not important. 
For concreteness, we adopt the unweighted mean of the four tomographic bins' SOMPZ from DES Y3 \citep{Myles2021}.
We apply our deprojection method to each individual shear profile, rescale by $r_{200c}$, and then stack all shear profiles using weights proportional to $r^2$.
The $r^2$ weights are a simple proxy for the inverse variance weights we use in our observational results, since these roughly scale as the angular area of a given radial bin, and our simulated clusters are all at the same redshift.

\begin{figure}
  \includegraphics[width=\columnwidth]{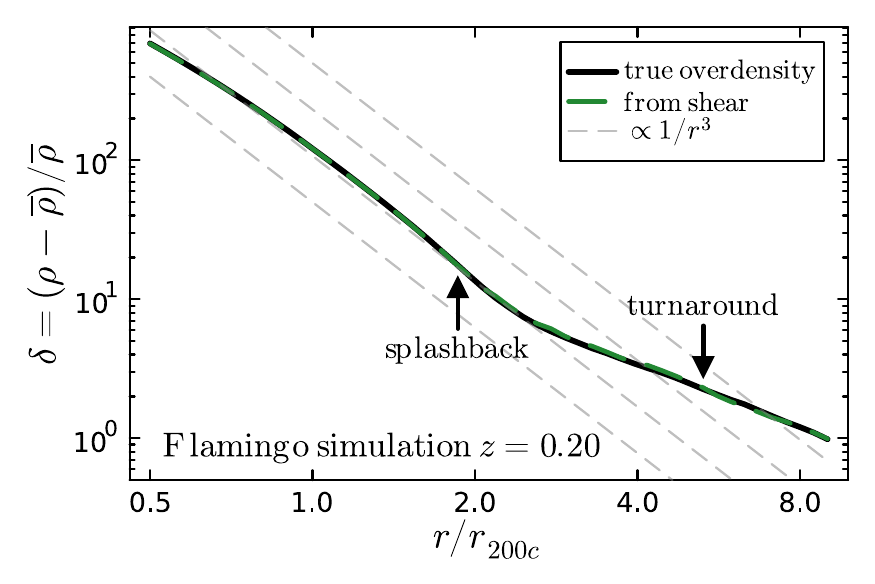}
  \caption{%
   Verification of our deprojection method using shear profiles from the Flamingo hydrodynamic simulations.
   We apply our deprojection formula Eq.~\eqref{eq:rho_from_ESD} to shear profiles generated from each simulated cluster by projecting along the simulation's three Cartesian axes.
   Then we stack the resulting overdensity profiles at fixed $r/r_{200c}$ (dashed green line).
   We successfully recover the true overdensity profile (solid black line) because the effects of non-symmetric mass distributions average out correctly when stacking many clusters.%
  }
  \label{fig:flamingo}
\end{figure}

\begin{figure}
  \includegraphics[width=\columnwidth]{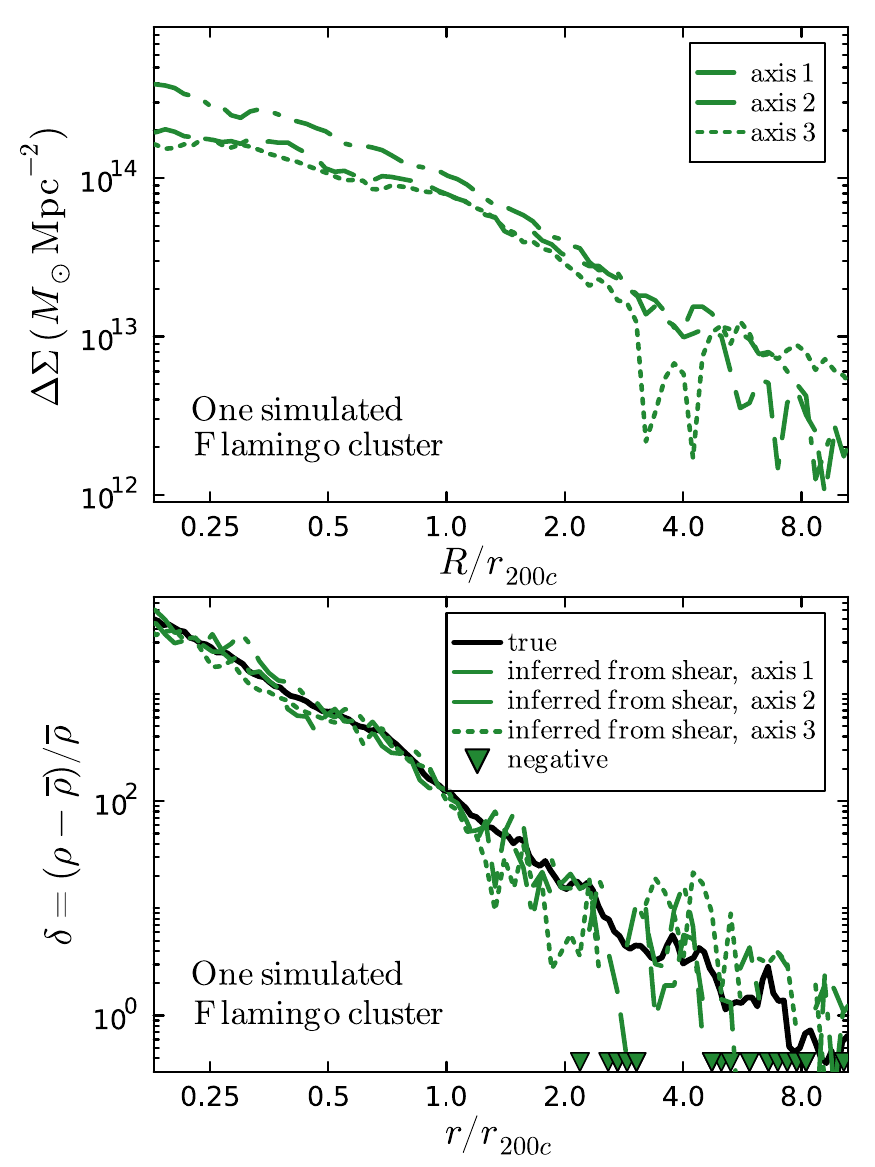}
  \caption{%
   \emph{Top}: Shear profiles of a single simulated galaxy cluster.
   We show three different profiles, obtained by projecting along the simulation's three Cartesian axes.
   They differ because the galaxy cluster and its environment are not symmetric.
   \emph{Bottom}: The density profiles inferred by our deprojection method using the shear profiles from the top panel.
   They again differ due to the non-symmetry of the cluster and its environment.
   The effects from non-symmetry average out, however, when stacking results from many different lines of sight (Fig.~\ref{fig:flamingo}).
  }
  \label{fig:flamingo_single}
\end{figure}

\begin{figure}
  \includegraphics[width=\columnwidth]{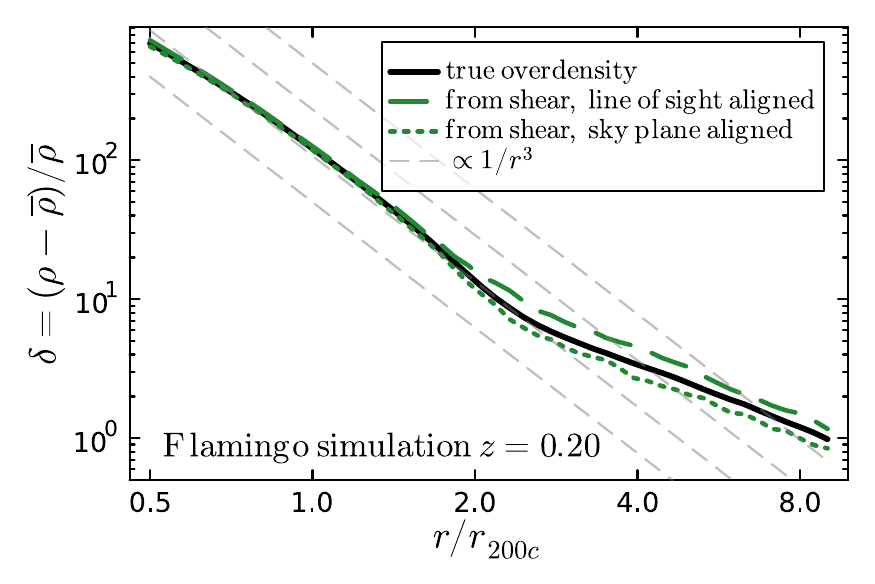}
  \caption{%
   Same as Fig.~\ref{fig:flamingo} but instead of projecting along all three Cartesian axes of the simulation we project each cluster along the Cartesian axis most aligned with its major axis (dashed line) and the axis least aligned with its major axis (dotted line).
   This illustrates results for clusters that are aligned with our line of sight or the sky plane.
  }
  \label{fig:flamingo_major}
\end{figure}

\begin{figure}
  \includegraphics[width=\columnwidth]{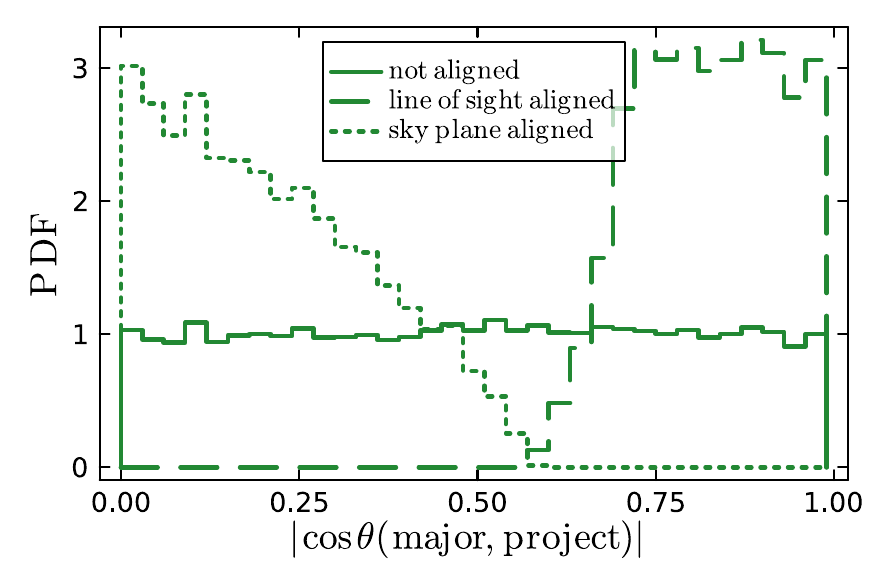}
  \caption{%
   The distribution of the cosine between the projection axis and each cluster's major axis for the different alignments shown in Fig.~\ref{fig:flamingo_major} (dashed and dotted lines).
   For reference, the unaligned case from Fig.~\ref{fig:flamingo} is also shown (solid line).
  }
  \label{fig:flamingo_angles}
\end{figure}

Our results are shown in Fig.~\ref{fig:flamingo}.
We find the expected behavior, namely an $r^{-3}$ decline in the one-halo region, with a slightly shallower slope at small radii, followed by a significantly shallower slope in the two-halo region at larger radii, where the surrounding structure becomes important.
For reference, Fig.~\ref{fig:flamingo} further shows the splashback and turnaround radii.
The splashback radius is located at $r/r_{200c} \approx 1.9$, towards the end of the one-halo region, marking the steepest slope in the stacked density profile.
The turnaround radius is located at $r/r_{200c} \approx 5.3$, far into the two-halo region.
Unlike in our observational results, we do not see a significant feature in the stacked overdensity profile at the turnaround radius.
Indeed, to our knowledge, no such feature is expected in $\Lambda$CDM, as we discuss in more detail in Sect.~\ref{sec:discussion} below.
For comparison, Fig.~\ref{fig:flamingo} also shows the true spherical overdensity profile, stacked in the same way as the one inferred from the shear profiles.
The true overdensity profile of each cluster is calculated by adding up all particles in spherical shells with bin edges that are evenly spaced in $\log_{10} r/\mathrm{Mpc}$ with width $0.01349$.
We see that the stacked deprojected overdensity profile matches the stacked true overdensity profile.
This agreement is possible because, for large samples of clusters, our deprojection formula recovers the true spherically averaged density even for non-symmetric lenses (see Sect.~\ref{sec:method:shear_to_rho}).

To further illustrate this point, Fig.~\ref{fig:flamingo_single} shows the shear profiles and the corresponding inferred density profiles for a single simulated cluster as observed along the three Cartesian axes.
The cluster and its environment are not symmetric, which leads to very different shear profiles and inferred density profiles for the three axes. Indeed, depending on the direction of the projection (the line of sight), filaments, voids, and other neighboring structures will be aligned in a different way and will affect the shear accordingly. This is especially relevant in the two-halo region.
Nonetheless, after stacking many clusters, we recover the true spherically averaged density profile (Fig.~\ref{fig:flamingo}).
When clusters are successively added to the stack, agreement with the true spherically averaged density profile is first achieved in the one-halo region.
The two-halo region requires more clusters to converge to the final result because of the higher degree of cluster-by-cluster variation in the outskirts.
These results show that our deprojection method successfully recovers the spherically averaged three-dimensional density profile, even in the absence of spherical symmetry.
This shows that the theoretical arguments from Sect.~\ref{sec:method:shear_to_rho} apply to realistic mass distributions.

The assumption underlying these results is that the clusters are randomly oriented with respect to our line of sight.
This is a reasonable assumption for the X-ray-selected eRASS:5 cosmology sample but it may be violated to some degree by our additional optical selection, i.e. by our $15 \leq \lambda < 100$ cut \citep[e.g,][]{Dietrich2014,Abbott2020,Wu2022,Sunayama2020}.
In particular, a richness selection may cause the halo's major axis to be aligned with our line of sight.
Figure~\ref{fig:flamingo_major} illustrates how samples of clusters that are aligned with our line of sight or the plane of the sky would affect our results.
In particular, we show results when projecting each cluster along the Cartesian axis closest to and furthest from its major axis.
This is likely a stronger alignment than that of our richness cuts (see Fig.~\ref{fig:flamingo_angles}), but it gives a useful illustration.
For the line-of-sight-aligned case, we find that our inferred density profiles are overestimated by about $5\%$ in the one-halo region and by about $25\%$ in the two-halo region.
For the sky plane aligned case, we find that our inferred density profiles are underestimated by about $5\%$ in the one-halo region and by about $15\%$ in the two-halo region.
In either case, we still roughly recover the correct slope in both regions, and, importantly, the inferred density profiles remain monotonically decaying, with no non-monotonic feature such as the one we see in our observational results at $x \sim 6$.
This indicates that the feature at $x \sim 6$ is unlikely to be due to optical projection effects.

The same conclusion about richness selection effects is also supported by the results of \citet{Sunayama2020}.
Indeed, we have applied our deprojection formula to their simulated shear profiles of optically selected clusters and found that optical selection does not induce a non-monotonic feature.
We have further found that, instead, a non-monotonic feature \emph{can} be induced by applying a sufficiently strong correction to the shear profiles that goes into the \emph{opposite} direction of what richness selection does according to \citet{Sunayama2020,Wu2022}.
This strengthens the conclusion that richness selection does not induce non-monotonic features.

\section{Robustness checks}
\label{sec:results:checks}

In this section, we assess the robustness of the feature at $x \sim 6$.
In particular, we consider median stacking, interpolation and extrapolation of the shear profiles, radius-dependent weights, the cross component of the shear, randoms, and running stacks.
Further checks can be found in Appendix~\ref{sec:appendix:more_robustness}, in particular regarding density profiles without rescaling, stacked shear, miscentering, the shear and selection responses, cluster member contamination, eRASS1, reduced shear, intrinsic alignment, the lens survey edges, scatter in our mass proxy, and a more precise $P(z_s)$.

\begin{figure}
  \includegraphics[width=\columnwidth]{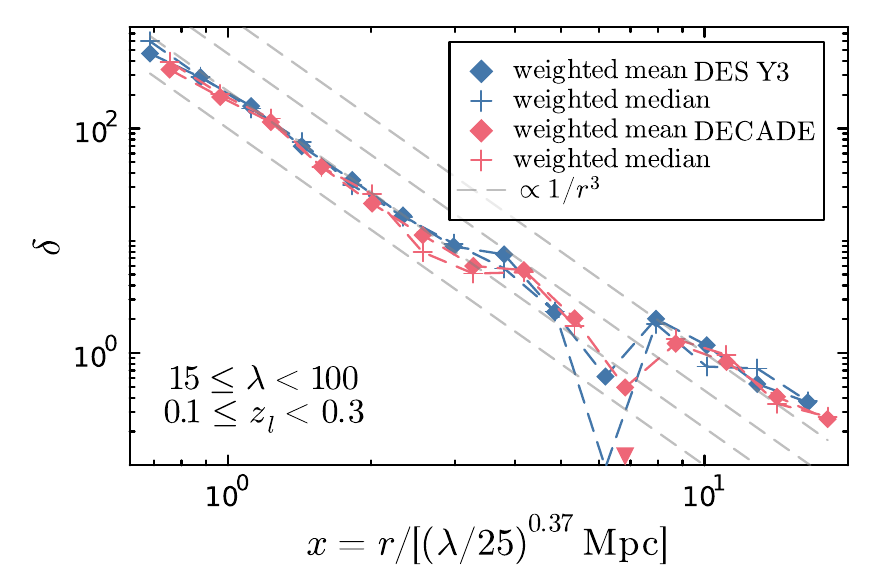}
  \caption{%
   Same as Fig.~\ref{fig:rho_of_x} but additionally showing the result when stacking using a weighted median (`+' symbols) instead of a weighted mean (diamond symbols).
   Error bars are left out for visual clarity.
   The feature at $x \sim 6$ is more pronounced with median stacking, indicating that it is not driven by outliers.
  }
  \label{fig:rho_of_x_median}
\end{figure}

\subsection{Median stacking}

Our stacked density profiles are obtained by taking a weighted mean of the density profiles of the individual clusters (see Sect.~\ref{sec:method:stacking}).
To assess robustness to outliers, we rerun our analysis using a weighted median instead.
Figure~\ref{fig:rho_of_x_median} shows that our results do not change significantly, except that the feature becomes even more pronounced.
This shows that our results, particularly the feature at $x \sim 6$, are not driven by outliers.

\begin{figure}
  \includegraphics[width=\columnwidth]{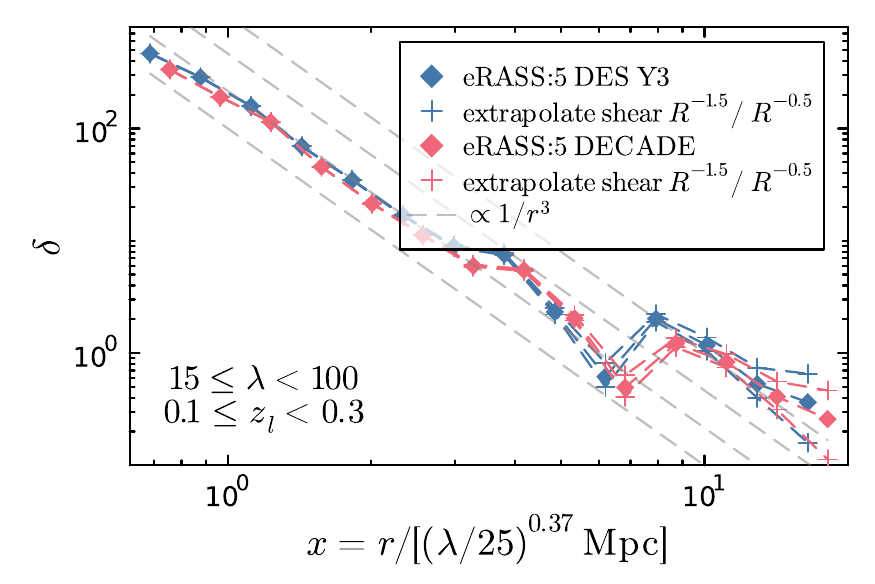}
  \caption{%
   Same as Fig.~\ref{fig:rho_of_x} but additionally showing the result for different choices of how to extrapolate beyond the outermost shear data point (Sect.~\ref{sec:method:shear_to_rho}).
   Our main analysis assumes a $R^{-1}$ decline (diamond symbols).
   The dashed lines assume $R^{-3/2}$ and $R^{-1/2}$ declines instead (`+' symbols).
   Error bars are left out for visual clarity.
   Extrapolation affects mainly the last few data points; the $x \sim 6$ feature is largely unaffected.%
  }
  \label{fig:rho_of_x_extrapolate}
\end{figure}

\subsection{Extrapolation and interpolation}

As discussed in Sect.~\ref{sec:method:shear_to_rho}, we interpolate and extrapolate the measured shear profiles when evaluating the integrals in the deprojection formula Eq.~\eqref{eq:rho_from_ESD}.
Extrapolation is an edge effect and becomes important only close to the outermost data point.
Our fiducial choice of extrapolation is to assume that the shear declines as $R^{-1}$ beyond the outermost shear data point.
This is close to the expected slope of the two-halo term in $\Lambda$CDM.
Figure~\ref{fig:rho_of_x_extrapolate} shows the results when extrapolating assuming $R^{-3/2}$ and $R^{-1/2}$ instead.
This range encompasses the plausible range in $\Lambda$CDM, and we have verified that it includes the two-halo term as calculated with CAMB \citep{Lewis2002}.
Figure~\ref{fig:rho_of_x_extrapolate} confirms that the choice of extrapolation has an effect only at large radii.
At the location of the $x \sim 6$ feature, the effect is insignificant.
This shows that the feature is not induced by our choice of extrapolation method for the shear profiles.

Interpolation is unimportant when the shear data points are close together.
Thus, interpolation plays a role mainly in our analysis using radial bins, not in our analysis using sliding windows.
Our fiducial choice is to interpolate quadratically.
With this choice of interpolation, the analyses using radial bins and using radial sliding windows agree well with each other.
We have rerun the analysis using linear interpolation and found that the difference is negligible for radial windows, while the feature becomes slightly less pronounced for radial bins.
This is expected since, as already mentioned in Sect.~\ref{sec:method:shear_to_rho}, mock tests have shown that the combination of wide radial bins and linear interpolation is inadequate for the purpose of recovering localized features.
The results with sliding windows and the agreement for quadratic interpolation show that the feature is not induced by our choice of interpolation.

\begin{figure}
  \includegraphics[width=\columnwidth]{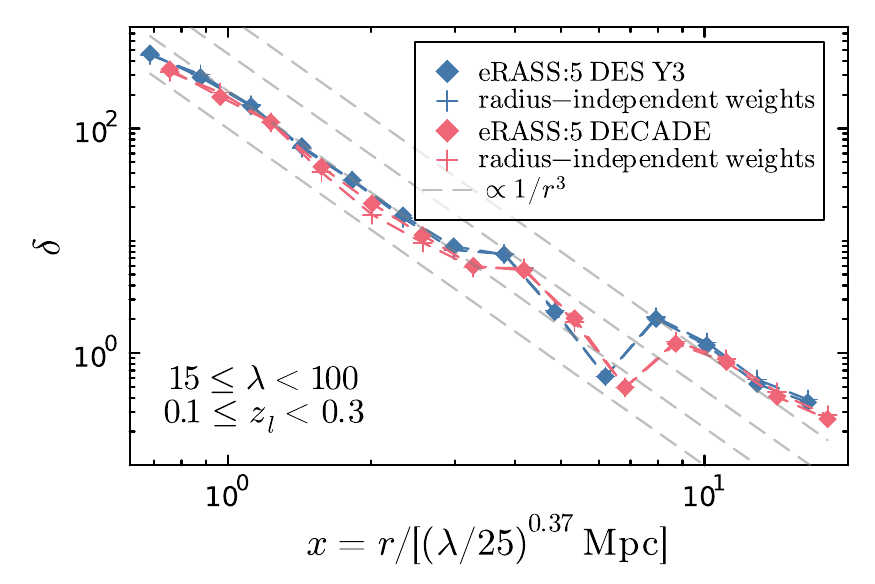}
  \caption{%
   Same as Fig.~\ref{fig:rho_of_x} but additionally showing the result when using the same stacking weights at all radii.
   In particular, we reuse the weights from the 10th radial bin for all other radial bins.
   Error bars are left out for visual clarity.
   Our results are mostly unaffected by whether or not radius-dependent weights are used.%
  }
  \label{fig:rho_of_x_weights}
\end{figure}

\subsection{Radius-dependent weights}

Our stacking weights $w_{l i}$ are determined separately for each radial bin or sliding window (see Sect.~\ref{sec:method:stacking}).
Thus, in principle, a radial trend in the stacked result may be due to a feature in the weights instead of one in the stacked quantities themselves.
In practice, however, this is not the case since the normalized weights are relatively constant as a function of radius.
Indeed, as Fig.~\ref{fig:rho_of_x_weights} shows, the results change very little if we re-run the analysis using the same weights at all radii.
Our results regarding stacked shear profiles from Appendix~\ref{sec:appendix:shear} lead to the same conclusion.

\begin{figure}
  \includegraphics[width=\columnwidth]{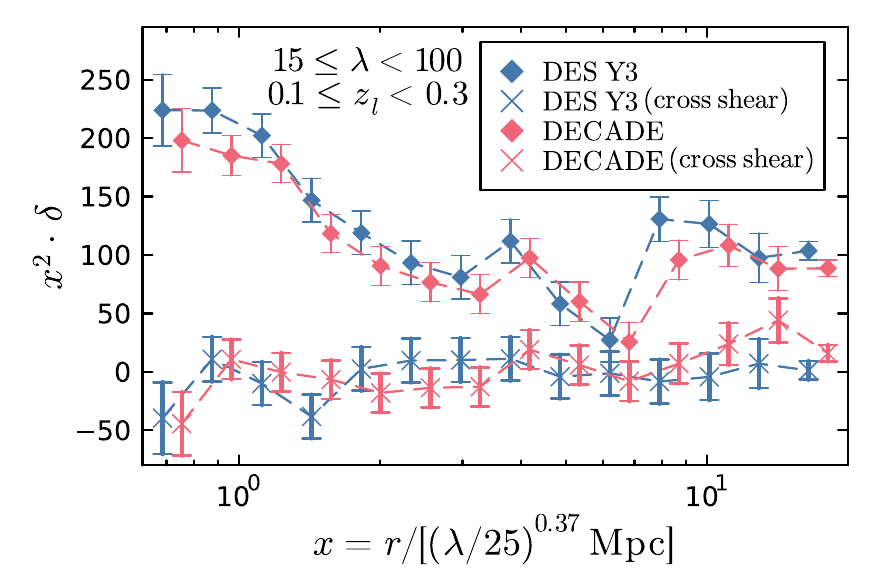}
  \caption{%
   Overdensity profiles using the cross (instead of the tangential) component of the shear multiplied by $x^2$ (cross symbols).
   The original result using the tangential shear is shown by the diamond symbols.
   The cross component results are consistent with zero.
  }
  \label{fig:cross_shear}
\end{figure}

\subsection{Cross shear}

For a pure lensing signal, the cross component of the shear vanishes.
Thus, we test for systematic effects from the instrument or data processing by re-running our analysis using the cross-component of the shear rather than the tangential component.
The result should be consistent with zero, and Fig.~\ref{fig:cross_shear} shows that this is indeed the case.
For simplicity, we do not correct for cluster-member contamination or subtract randoms.
The reduced $\chi^2$ values, calculated using the shape-noise-only covariance matrix, are $0.60$ and $1.18$ for DES~Y3 and DECADE, respectively.

\begin{figure}
  \includegraphics[width=\columnwidth]{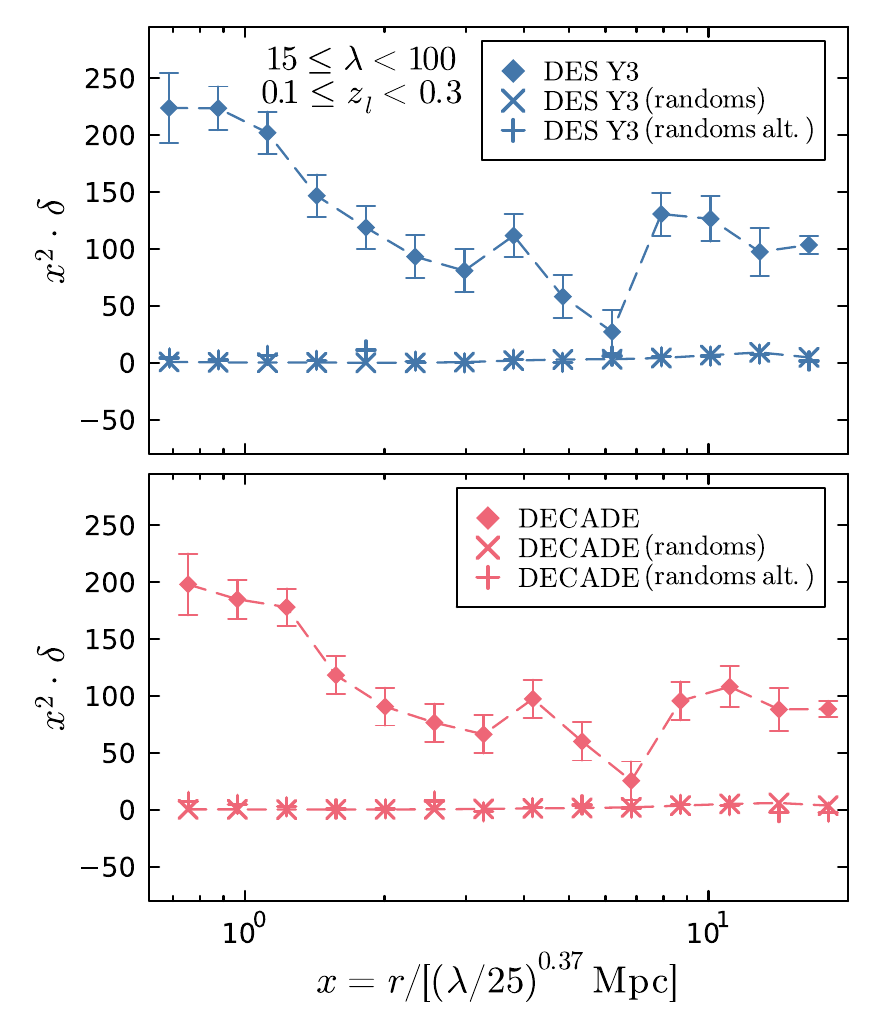}
  \caption{%
   The randoms' overdensity profiles we subtract from our results as described in Sect.~\ref{sec:method:randoms} (cross symbols), multiplied by $x^2$.
   For reference, the original (unsubtracted) result is shown by the diamond symbols.
   The `+' symbols show the results for randoms obtained using an alternative procedure based on a preliminary version of the eRASS:5 selection function from the eROSITA digital twin \citep{Seppi2022,erass5catalog}.
  }
  \label{fig:rho_of_x_randoms}
\end{figure}

\subsection{Randoms}
\label{sec:results:checks:randoms}

Figure~\ref{fig:rho_of_x_randoms} shows the randoms we subtract as described in Sect.~\ref{sec:method:randoms}.
For simplicity, we do not correct for cluster-member contamination here.
We find that randoms have only a minor effect in both DES Y3 and DECADE.
As a further cross-check, Fig.~\ref{fig:rho_of_x_randoms} also shows randoms generated using an alternative procedure that differs from the \Healpix-based subtraction from Sect.~\ref{sec:method:randoms}.
Namely, we use a preliminary version (`v3') of the selection function of the eRASS:5 cosmology sample, as obtained from the eROSITA digital twin \citep{Seppi2022,erass5catalog}, to generate a sample of redshifts, coordinates, and count rates that is 30 times as large as the full eRASS:5 cosmology sample.
Since these randoms do not include richness, we use the eRASS1 count-rate and richness-mass scaling relations \citep{Ghirardini2024} to derive richness values.
We then re-run our analysis on this random catalog.
The results are consistent with our main random subtraction procedure, implying only a negligible effect.

\begin{figure*}
  \centering
  \includegraphics[width=1.8\columnwidth]{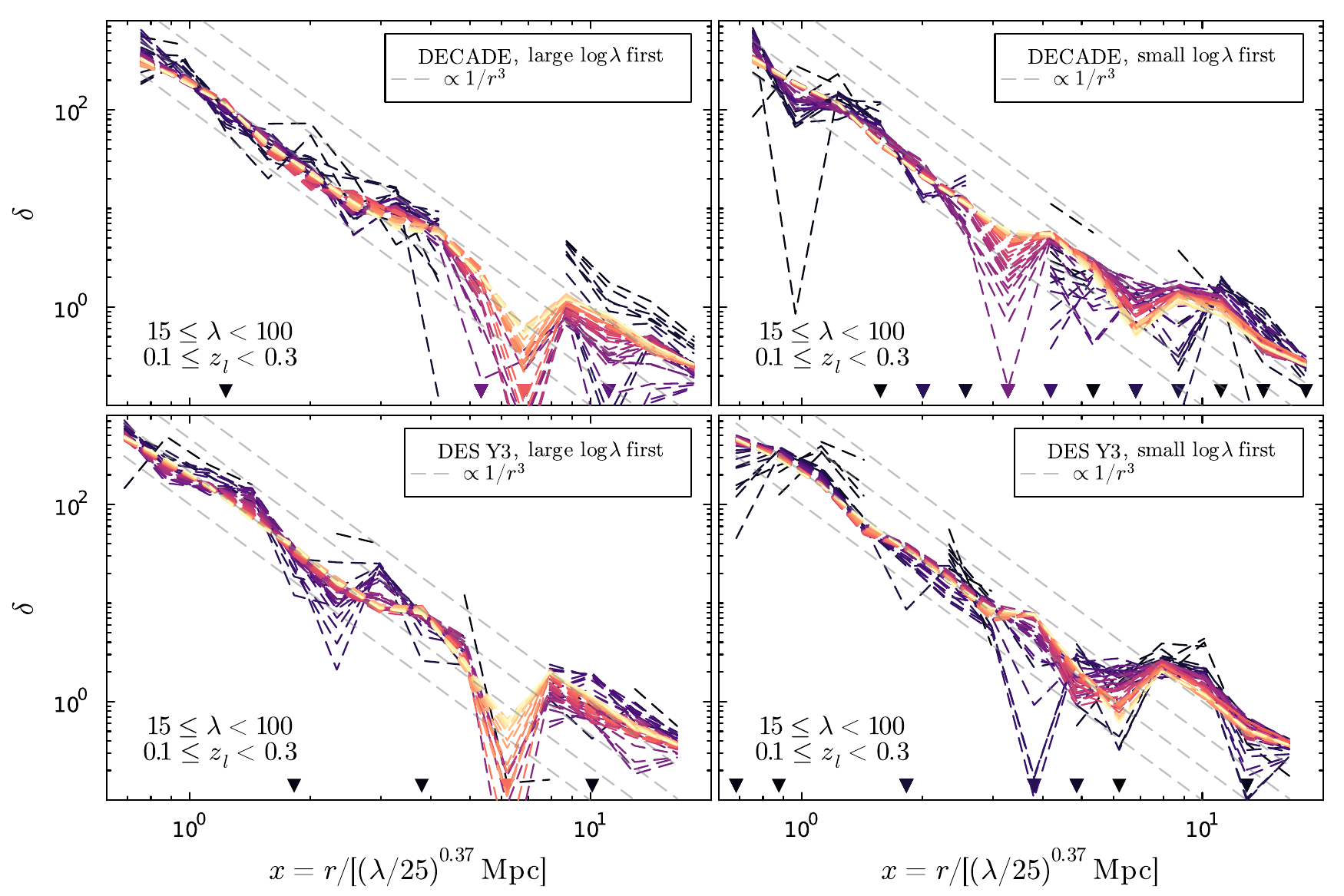}
  \caption{%
   Running stacks of the overdensity profiles for DES Y3 (bottom) and DECADE (top).
   That is, we sort the clusters by richness, then split the clusters into 70 batches, plotting the intermediate result after adding each batch to the stack.
   The colors are brighter the more clusters are added to the stack.
   The final step (brightest color) matches the result from Fig.~\ref{fig:rho_of_x}.
   We show results when adding clusters in order of decreasing (left) and increasing (right) richness.
   If a batch contains fewer than five clusters, we combine it with the following batch.
   Triangles indicate negative values.
   The results indicate that the feature is strongest at high richness.
  }
  \label{fig:step_by_step_lambda}
\end{figure*}

\begin{figure*}
  \centering
  \includegraphics[width=1.8\columnwidth]{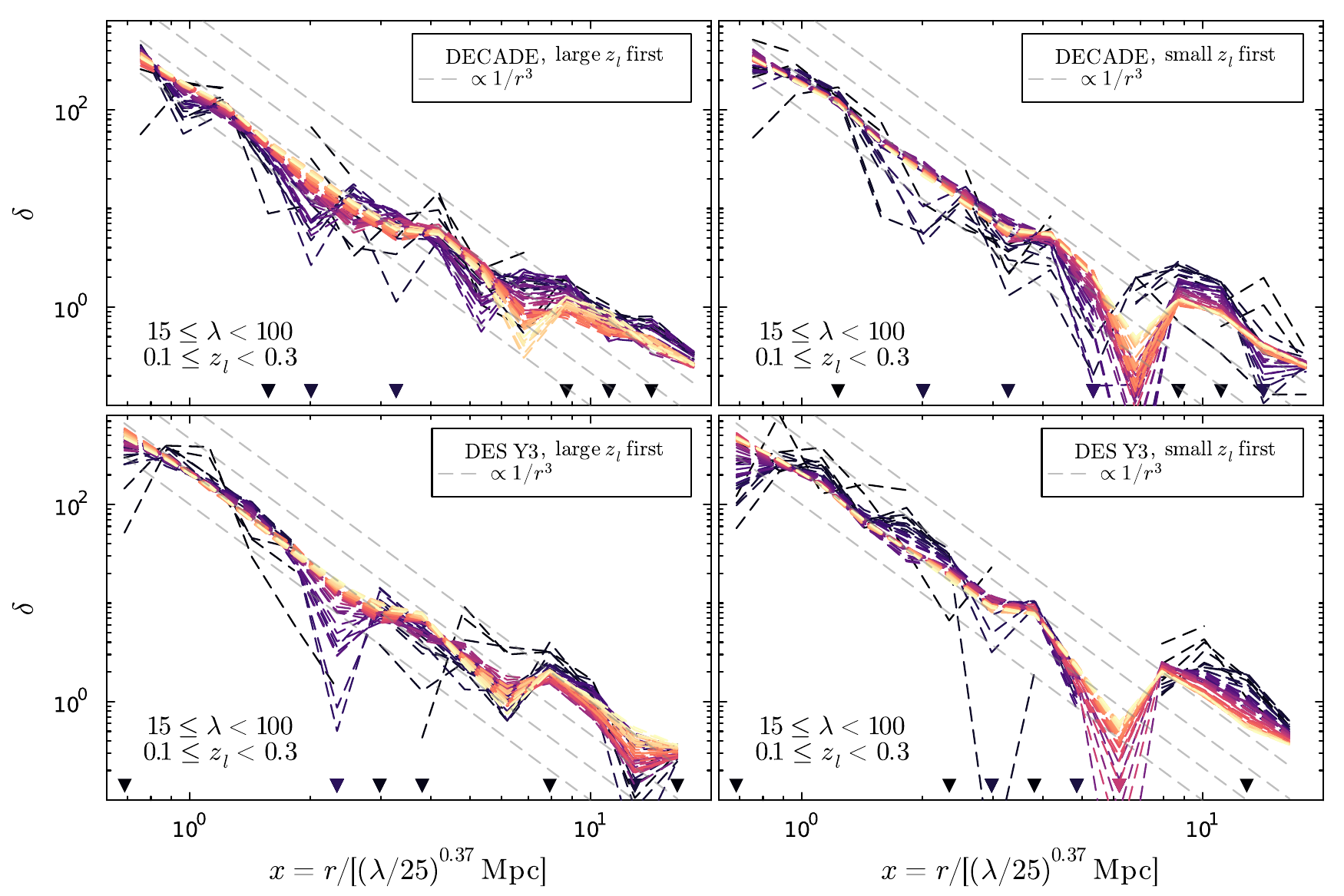}
  \caption{%
   Same as Fig.~\ref{fig:step_by_step_lambda} but sorting the clusters by lens redshift $z_l$ instead of $\log \lambda$.
   The results indicate that the feature is strongest at low redshift.
  }
  \label{fig:step_by_step_z}
\end{figure*}

\subsection{Running stacks}

Figures~\ref{fig:step_by_step_lambda} and \ref{fig:step_by_step_z} illustrate how our stacked density profiles are built from the individual clusters:
Instead of showing only the final stack, these figures also show intermediate results after having added small batches of clusters to the stack.
In particular, we first sort the clusters by richness or redshift and then add them to the stack in batches, plotting the result after each batch.
Figure~\ref{fig:step_by_step_lambda} adds clusters in order of increasing and decreasing $\log \lambda$.
Figure~\ref{fig:step_by_step_z} adds clusters in order of increasing and decreasing $z_l$.
These results suggest that the feature is strongest at high richness and low redshift.
Indeed, when adding the high-richness clusters first, the feature around $x \sim 6$ is strong and becomes somewhat weaker when the lower-richness clusters are added.
In contrast, with only the low-richness clusters added, the feature is less pronounced and builds up when adding the higher-richness clusters.
We see analogous behavior when adding the clusters sorted by redshift.
This is consistent with the results from Sect.~\ref{sec:results}, where we split the sample into low/high richness and low/high redshift.
At low richness and high redshift, where the feature is less pronounced, there are hints of a secondary feature around $x \sim 2 - 3$, consistent with what we saw in Fig.~\ref{fig:rho_of_x_lowlambda} and Fig.~\ref{fig:rho_x_MrescaledBulbul19}, although this might just be noise and, in the case at hand, goes away in the final stack.
We also see that, overall, the feature at $x \sim 6$ remains at a consistent location and exhibits no sudden jumps, smoothly converging to the final result as the number of stacked clusters increases, consistent with it being a real feature rather than noise.

\section{Discussion}
\label{sec:discussion}

In this section, we discuss the interpretation and implications of our results from Sect.~\ref{sec:results}.
In particular, we discuss the extended $r^{-3}$ decline, how it constrains modified gravity, the splashback radius, and possible origins of the feature at $x \sim 6$ (where $x$ is the radius rescaled by our observational proxy for cluster size).

\subsection{Direct confirmation of $r^{-3}$ decline}

In the one-halo region, our density profiles qualitatively match expectations from $\Lambda$CDM.
In fact, for the first time, we directly confirm the $r^{-3}$ decline predicted by numerical simulations of cold dark matter halos \citep{Navarro1995, Navarro1996}.
Many previous studies have already found agreement between NFW halos and observations.
However, these usually either perform a parametric fit and thus are not direct measurements, or operate at the level of the mass profile rather than the density profile \citep[e.g.][]{Umetsu2025}.
To the best of our knowledge, the only previous comparable measurements of the density profiles in galaxy clusters were done in \citet{Johnston2007b} and \citet{Mistele2025}.
They analyzed stacked SDSS clusters \citep{Koester2007} and the individual clusters from the CLASH sample \citep{Postman2012}, respectively.
Their results were also consistent with a $r^{-3}$ decline but with much larger error bars, which precluded a definitive conclusion.
Due to the statistical power of the eRASS:5 galaxy cluster sample, we obtain a significantly stronger result here.

\subsection{Constraints on modified gravity}

Our finding of an extended region with a $r^{-3}$ decline also constrains modified gravity theories.
For example, it has been a long-standing issue for theories based on Modified Newtonian Dynamics \citep[MOND,][]{Milgrom1983a,Milgrom1983c} that galaxy clusters do not follow the same scaling relations as galaxies \citep[e.g.][]{Sanders1999,Sanders2003,Eckert2022}; in particular clusters do not follow the Baryonic Tully-Fisher Relation \citep[BTFR][]{McGaugh2000} and the Radial Acceleration Relation \citep[RAR][]{Lelli2017b,Brouwer2021,Mistele2023d}.
One proposed solution is that there is missing mass, i.e., mass that is so far unobserved, in the centers of galaxy clusters.

Our results show that there is more to explain (see also Appendix~\ref{sec:appendix:MOND}).
Indeed, the slope in the outskirts of an isolated cluster in MOND is $-2$, which differs significantly from our results, which are close to $-3$.
However, clusters are not isolated, and so the external field effect may play a role \citep[e.g.][]{Milgrom2008,Kelleher2024,Famaey2024}.
In particular, for a constant external field, we expect a transition to $\rho(r) = (5/6) (a_0/g_e) \rho_b (r)$, where $a_0$ is the MOND acceleration constant, $g_e$ is the external field, and $\rho_b$ is the baryonic mass density.\footnote{
  This formula can be derived using standard techniques in the deep-MOND limit of QUMOND, assuming a spherically symmetric baryonic density profile and a constant external field.
  This is in contrast to point masses, which exhibit a steep slope of $-7$ \citep{Bilek2024}.
}
The hot gas around galaxy clusters, which dominates $\rho_b(r)$, may have a slope close to $-3$ over at least some radial range \citep[e.g.,][]{Zhang2025}, but whether such an explanation is feasible remains to be seen.
A full explanation might also need to include a more realistic external field, nearby structures, or effects from averaging over clusters with a range of properties and environments.

Our results are further relevant to fully relativistic models such as Aether Scalar Tensor Theory \citep[AeST,][]{Skordis2020}\footnote{
 More generally, our results can be reinterpreted in any minimally coupled metric theory in which the metric can be written as an FLRW background with weak-field potentials on top.
 Our overdensity profiles then correspond to $(4 \pi G_N)^{-1} \Delta \phi_W$ where $\phi_W$ is the Weyl potential.
} which can be viewed as hybrid models \citep{Mistele2023d} that interpolate between MOND-like behavior on galactic scales and $\Lambda$CDM-like behavior on cosmological scales.
Galaxy clusters sit right in the middle between these opposites and exhibit a rich phenomenology in AeST, including oscillations of the gravitational force \citep[e.g.,][]{Durakovic2023,Mistele2023d}.
This oscillatory behavior is reminiscent of the feature we see in the measured density profiles, highlighting another potential way for MOND to accommodate our results.

\subsection{Splashback radius and two-halo region}

Perhaps the most well-known location in the outskirts of galaxy clusters is the splashback radius, whose characteristic property is a drop in density corresponding to accreting collisionless matter reaching its apocenter after first infall \citep{Diemer2014,Adhikari2014}.
The splashback radius is located close to $r_{200m}$, which roughly corresponds to $x \sim 1.7$ in terms of our rescaled radius $x$ (see Fig.~\ref{fig:Delta_of_x}).
From a visual inspection of our non-parametric density profiles in Fig.~\ref{fig:rho_of_x}, we do not find a strong steepening at this location.
Our analysis using sliding windows in Fig.~\ref{fig:rho_of_x_windows} does, however, show a slight drop in density at this location.
The effect is small and may be due to noise, but our results are consistent with a splashback feature as expected in $\Lambda$CDM, since the expected steepening is only modest in stacked density profiles.
Further, from parametric fits to our non-parametric density profiles, we found the location of the steepest slope to be roughly compatible with the expected $x \sim 1.7$.
At larger radii, around $x \sim 3$, we see a flattening of the density profiles, which signals the transition to the two-halo region.
Apart from the feature at $x \sim 6$ (discussed in the next subsection), our results for the two-halo region are compatible with expectations from $\Lambda$CDM and earlier lensing studies \citep[e.g.,][see also Fig.~\ref{fig:literature}]{Umetsu2017,Chang2018}.

\subsection{The feature}

In the context of $\Lambda$CDM, the most surprising aspect of our results is the feature at $x \sim 6$.
In Sect.~\ref{sec:results:checks}, we showed that the feature survives many robustness checks.
Here, we discuss its interpretation, assuming it's a real physical effect.
We currently lack a clear understanding of its origin, but it is still worthwhile to discuss possible relations and connections for future investigation.

The feature's location, and that location's scaling with cluster mass, are consistent with the turnaround radius.
However, at least in $\Lambda$CDM, cosmological simulations are not seeing a non-monotonic feature at the turnaround radius \citep[e.g.,][]{Korkidis2025}.
Another relevant location is the depletion radius \citep{Fong2021,Fong2022,Han2026}, where mass accretion onto neighboring halos, competing for the remaining matter in between, can lead to relative voids.
\citet{Fong2021} show that this can lead to a non-monotonic feature in the density profile, but mainly in low-mass halos.
This differs from the $x \sim 6$ feature we observe, which is strongest at high richness.
One might further consider that clusters are located near voids, which lie between the filaments that connect the clusters to the cosmic web.
In that view, the trough-like shape of our feature may be due to the lower void densities.
This is complementary to the possibility that our feature is related to the filaments themselves.
Indeed, \citet{Sen2026} find that, in simulations, density profiles along individual directions can be non-monotonic.
However, the stacked, spherically averaged profile declines smoothly.
Thus, in our observational results, we also expect a smooth decline.
This assumes that our lines of sight are randomly oriented with respect to the filaments.
If this is not the case, a more complex stacked profile may be possible, although our results from Sect.~\ref{sec:method:hydro} regarding specially oriented halos do not support such a scenario.
In any case, such scenarios may be tested by cross-correlating the eRASS:5 clusters with tracers of the cosmic web and large-scale structure.

More generally, our results can be affected by selection effects.
However, we find a similar feature in the SZ-selected SPT-3G sample and when replacing our richness cuts by X-ray based cuts.
Furthermore, we find consistent results for DES and DECADE despite the fact that the eRASS selection function varies across the sky and, in particular, differs significantly between the DES~Y3 and DECADE footprints \citep{Clerc2024,Ghirardini2024}.
This suggests that the feature's existence is robust to selection effects.
Whether its precise location and shape depend on sample selection remains to be studied in more detail in future work.

The feature is at least superficially reminiscent of those detected in the SZ profiles from \citet{Anbajagane2022,Anbajagane2024}.
Since the SZ effect measures gas pressure, these may be explained by gas shocks.
In contrast, our weak-lensing analysis is sensitive to the total mass density, which is expected to be dominated by dark matter.
In $\Lambda$CDM, the dark matter component is collisionless, so it is not expected to exhibit shocks.
Still, the gas and dark matter components are coupled gravitationally, meaning that a feature in one may, in principle, translate into a feature in the other.
In $\Lambda$CDM, however, we do not expect this to be a sufficiently large effect \citep[e.g.][]{Aung2021}.

\section{Conclusion}
\label{sec:conclusion}

In this study, we have measured stacked mass density profiles in the outskirts of the eRASS:5 galaxy clusters using weak-lensing data from DES Y3 and DECADE.
We provide the first direct, non-parametric measurement of spherically averaged three-dimensional density profiles with significant constraining power.
We constrain their radial shape and search for features in the infall region around the galaxy clusters.

We have separately measured density profiles for the clusters in the DES and DECADE footprints.
We find an extended radial region with a $r^{-3}$ decline for both DES and DECADE.
This confirms $\Lambda$CDM predictions for the one-halo term and has implications for alternative models of dark matter and modified gravity.
In the two-halo region, we find, for the first time, a non-monotonic feature roughly at the location of the turnaround radius.
The feature appears consistently in both DES and DECADE, and its location scales with cluster size.
Indeed, the feature becomes visible only when rescaling cluster-centric distances by an observational proxy of a cluster's radial extent.
The feature passes extensive robustness checks and, statistically, a simple broken power law is ruled out at more than $5\sigma$.
We do not detect the feature at high redshift, $z_l \geq 0.3$, but we do see it consistently at both high and low richness.
More generally, the feature is strongest at high richness and low redshift.

The physical interpretation of the feature remains open, but its location at the turnaround radius and its scaling with cluster size suggest possible links to the dynamics of the outer infall region, dark matter, and the cosmic expansion history.
Further investigations of potential explanations are therefore crucial, both within $\Lambda$CDM and considering alternative models.
A full understanding of the density profiles in the infall region would also enable analyses using cluster outskirts to constrain structure formation and cosmology \citep{Mpetha2024,Mpetha2025}.
On the observational side, upcoming weak-lensing data sets from Stage-IV surveys such as Euclid, LSST, and Roman will probe important observational signatures, including redshift trends.
They will also be able to further scrutinize the statistical and systematic uncertainties of our measurements.

Our results demonstrate the power of direct, non-parametric measurements of galaxy cluster density profiles for detecting localized features in their outskirts.
This allows going well beyond the capabilities of simple parametric models and opens up a new window for studying galaxy clusters and their surrounding large-scale environment.

\begin{acknowledgements}
The authors thank Chiway Chang for the helpful discussions.
X. Zhang, E. Bulbul, E. Artis, and S. Zelmer acknowledge financial support from the European Research Council (ERC) Consolidator Grant under the European Union’s Horizon 2020 research and innovation program (grant agreement CoG DarkQuest No 101002585).
This work was performed in part at Aspen Center for Physics, which is supported by National Science Foundation grant PHY-2210452.
\\
This work is based on data from eROSITA, the soft X-ray instrument aboard SRG, a joint Russian-German science mission supported by the Russian Space Agency (Roskosmos), in the interests of the Russian Academy of Sciences represented by its Space Research Institute (IKI), and the Deutsches Zentrum f{\"{u}}r Luft und Raumfahrt (DLR). The SRG spacecraft was built by Lavochkin Association (NPOL) and its subcontractors and is operated by NPOL with support from the Max Planck Institute for Extraterrestrial Physics (MPE).

\\
The development and construction of the eROSITA X-ray instrument were led by MPE, with contributions from the Dr. Karl Remeis Observatory Bamberg \& ECAP (FAU Erlangen-Nuernberg), the University of Hamburg Observatory, the Leibniz Institute for Astrophysics Potsdam (AIP), and the Institute for Astronomy and Astrophysics of the University of T{\"{u}}bingen, with the support of DLR and the Max Planck Society. The Argelander Institute for Astronomy of the University of Bonn and the Ludwig Maximilians Universit{\"{a}}t Munich also participated in the science preparation for eROSITA.

\\
This project used public archival data from the Dark Energy Survey (DES). Funding for the DES Projects has been provided by the U.S. Department of Energy, the U.S. National Science Foundation, the Ministry of Science and Education of Spain, the Science and Technology FacilitiesCouncil of the United Kingdom, the Higher Education Funding Council for England, the National Center for Supercomputing Applications at the University of Illinois at Urbana-Champaign, the Kavli Institute of Cosmological Physics at the University of Chicago, the Center for Cosmology and Astro-Particle Physics at the Ohio State University, the Mitchell Institute for Fundamental Physics and Astronomy at Texas A\&M University, Financiadora de Estudos e Projetos, Funda{\c c}{\~a}o Carlos Chagas Filho de Amparo {\`a} Pesquisa do Estado do Rio de Janeiro, Conselho Nacional de Desenvolvimento Cient{\'i}fico e Tecnol{\'o}gico and the Minist{\'e}rio da Ci{\^e}ncia, Tecnologia e Inova{\c c}{\~a}o, the Deutsche Forschungsgemeinschaft, and the Collaborating Institutions in the Dark Energy Survey.

The Collaborating Institutions are Argonne National Laboratory, the University of California at Santa Cruz, the University of Cambridge, Centro de Investigaciones Energ{\'e}ticas, Medioambientales y Tecnol{\'o}gicas-Madrid, the University of Chicago, University College London, the DES-Brazil Consortium, the University of Edinburgh, the Eidgen{\"o}ssische Technische Hochschule (ETH) Z{\"u}rich,  Fermi National Accelerator Laboratory, the University of Illinois at Urbana-Champaign, the Institut de Ci{\`e}ncies de l'Espai (IEEC/CSIC), the Institut de F{\'i}sica d'Altes Energies, Lawrence Berkeley National Laboratory, the Ludwig-Maximilians Universit{\"a}t M{\"u}nchen and the associated Excellence Cluster Universe, the University of Michigan, the National Optical Astronomy Observatory, the University of Nottingham, The Ohio State University, the OzDES Membership Consortium, the University of Pennsylvania, the University of Portsmouth, SLAC National Accelerator Laboratory, Stanford University, the University of Sussex, and Texas A\&M University.
\\
Based in part on observations at Cerro Tololo Inter-American Observatory, National Optical Astronomy Observatory, which is operated by the Association of Universities for Research in Astronomy (AURA) under a cooperative agreement with the National Science Foundation.
\\
The Legacy Surveys consist of three individual and complementary projects: the Dark Energy Camera Legacy Survey (DECaLS; Proposal ID \#2014B-0404; PIs: David Schlegel and Arjun Dey), the Beijing-Arizona Sky Survey (BASS; NOAO Prop. ID \#2015A-0801; PIs: Zhou Xu and Xiaohui Fan), and the Mayall z-band Legacy Survey (MzLS; Prop. ID \#2016A-0453; PI: Arjun Dey). DECaLS, BASS and MzLS together include data obtained, respectively, at the Blanco telescope, Cerro Tololo Inter-American Observatory, NSF’s NOIRLab; the Bok telescope, Steward Observatory, University of Arizona; and the Mayall telescope, Kitt Peak National Observatory, NOIRLab. Pipeline processing and analyses of the data were supported by NOIRLab and the Lawrence Berkeley National Laboratory (LBNL). The Legacy Surveys project is honored to be permitted to conduct astronomical research on Iolkam Du’ag (Kitt Peak), a mountain with particular significance to the Tohono O’odham Nation.
\end{acknowledgements}

\bibliographystyle{aa}
\bibliography{references}

\begin{appendix}
\nolinenumbers %

\section{Further robustness checks}
\label{sec:appendix:more_robustness}

In this appendix, we present further robustness checks regarding density profiles without rescaling, stacked shear, miscentering, the shear and selection responses, cluster member contamination, eRASS1, reduced shear, intrinsic alignment, the lens survey edges, scatter in the mass proxy, and a more precise $P(z_s)$.

\begin{figure}
  \includegraphics[width=\columnwidth]{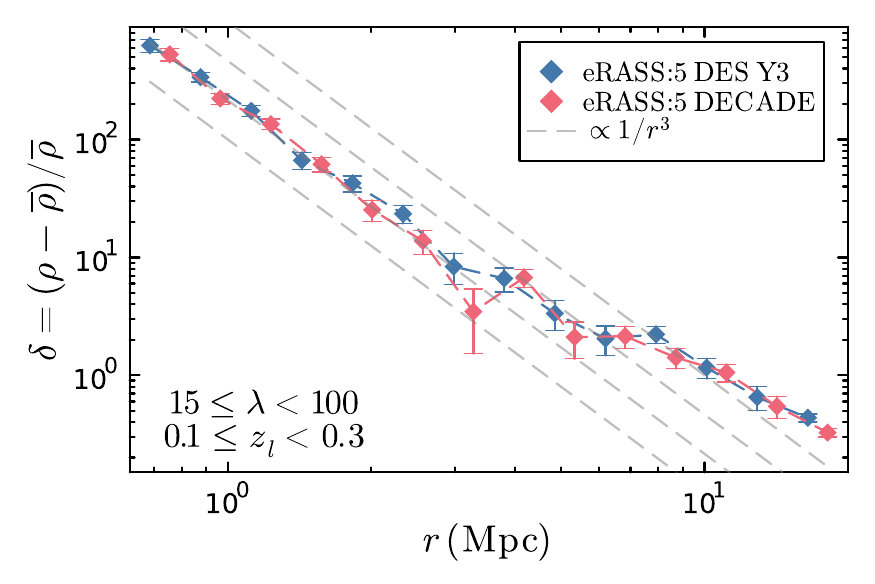}
  \caption{%
   Same as Fig.~\ref{fig:rho_of_x} but without rescaling the radius by cluster size.
   We still clearly see the one-halo term with its $r^{-3}$ decline followed by the shallower two-halo term, but there is no longer a clear feature.
   This indicates that the feature scales with cluster size and is washed out here. 
  }
  \label{fig:rho_of_r}
\end{figure}

\subsection{Density profiles without rescaling}

The importance of re-scaling cluster-centric distances by cluster size can be seen from Fig.~\ref{fig:rho_of_r}, which shows results \emph{without} such a rescaling.
We still clearly see the one-halo term with a $r^{-3}$ decline followed by a shallower two-halo term.
We, however, no longer see a clear feature consistent between DES and DECADE; we only see some complexity in the transition region.
This reinforces the point that the location of the feature scales with cluster size, since that explains why the feature gets washed out when stacking in $r$ but becomes visible when stacking in $x$.

\subsection{Stacked shear}
\label{sec:appendix:shear}

\begin{figure}
  \includegraphics[width=\columnwidth]{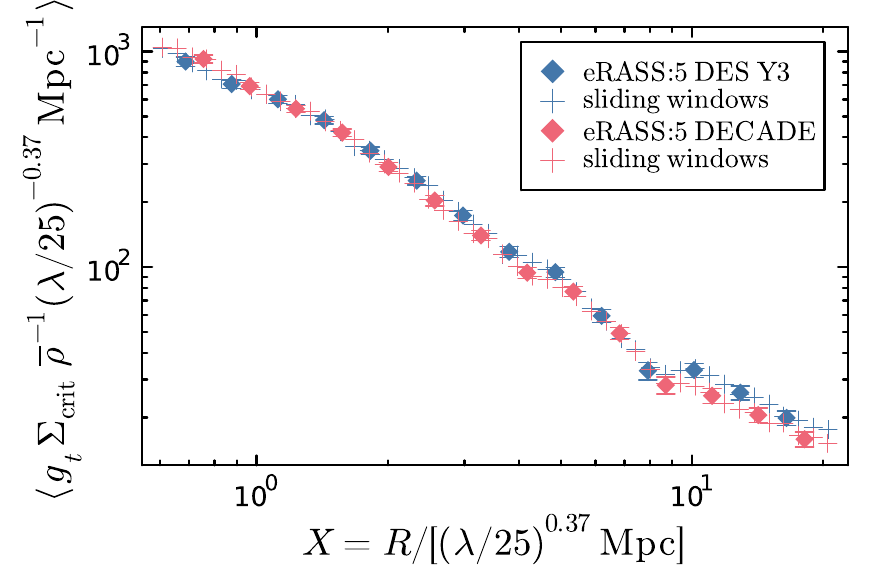}
  \includegraphics[width=\columnwidth]{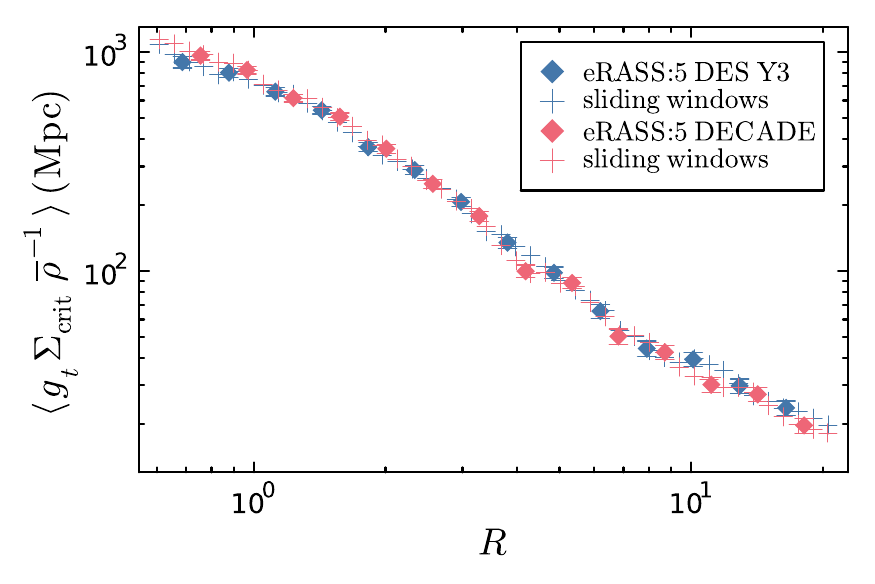}
  \caption{%
   Stacked shear profiles for the same cluster samples as the stacked overdensity profiles from Fig.~\ref{fig:rho_of_x_windows} and Fig.~\ref{fig:rho_of_x}.
   They are calculated using somewhat non-standard weights, as explained in the text.
   \emph{Top}: Stacked shear as a function of rescaled projected radius $X$. The stacked quantity includes the inverse background density around each lens and the inverse rescaling radius of each lens, as also explained in the text.
   \emph{Bottom}: Stacked shear as a function of projected radius $R$.
  }
  \label{fig:shear}
\end{figure}

The $x \sim 6$ feature in the stacked overdensity profiles has an analog in the corresponding stacked shear profiles.
Indeed, Fig.~\ref{fig:shear} (top panel) shows a shear profile stacked in terms of the re-scaled projected radius $X = R / [(\lambda/25)^{0.37}\,\mathrm{Mpc}]$ and we see a feature at $X \sim 7-10$.
As in the case of the density profiles, the feature goes away when stacking in terms of the original, not rescaled, radius $R$ (bottom panel).
That said, the feature is more subtle than in our overdensity profiles.
The reason is that shear is a projected and, partially, a cumulative quantity, which  leads to localized features in 3D density profiles being smeared out.

Given reasonable assumptions, the $x \sim 6$ feature in the stacked density profiles can be fully recovered from the stacked shear profiles.
In particular, the assumption is that the weights $w_{l i}$ we use to stack the overdensity profiles of the individual clusters are the same in each radial bin, $w_{l i} = w_l$.
Then, the stacked overdensity profile can be recovered by first obtaining a stacked shear profile as a function of $X$ (namely $\langle g_t \Sigma_{\mathrm{crit}} \overline{\rho}^{-1} (\lambda/25)^{-0.37} \,\mathrm{Mpc}^{-1} \rangle$ stacked with the weights $w_l$) and then applying the deprojection formula Eq.~\eqref{eq:rho_from_ESD}.
This is a somewhat non-standard stacked shear but it is the correct one to recover the overdensity profile.
 This can be shown mathematically by plugging the stacked shear into our deprojection formula and reordering the sum from the stack with the integral from the deprojection.
For this reason, the shear profiles in Fig.~\ref{fig:shear} re-use the stacking weights $w_{li}$ that we used for the overdensity profiles.
Moreover, we use the same weight for all radii (namely, for the binned analysis, the weight $w_{li}$ from radial bin $10$ and, for the analysis using sliding windows, the weight from sliding window $31$).
We have verified explicitly that applying our deprojection formula to the stacked shear profiles shown in the top panel of Fig.~\ref{fig:shear} recovers the stacked overdensity profiles from Fig.~\ref{fig:rho_of_x} to an excellent approximation.
This shows explicitly that our density feature is present but smeared out in the shear profiles.
It further shows that it makes not much of a difference whether we deproject first and then stack (as done in the main text) or stack first and then deproject (as done here).
And it shows that the radial dependence in the weights $w_{l i}$ is unimportant (see also Sect.~\ref{sec:results:checks}).

\begin{figure}
  \includegraphics[width=\columnwidth]{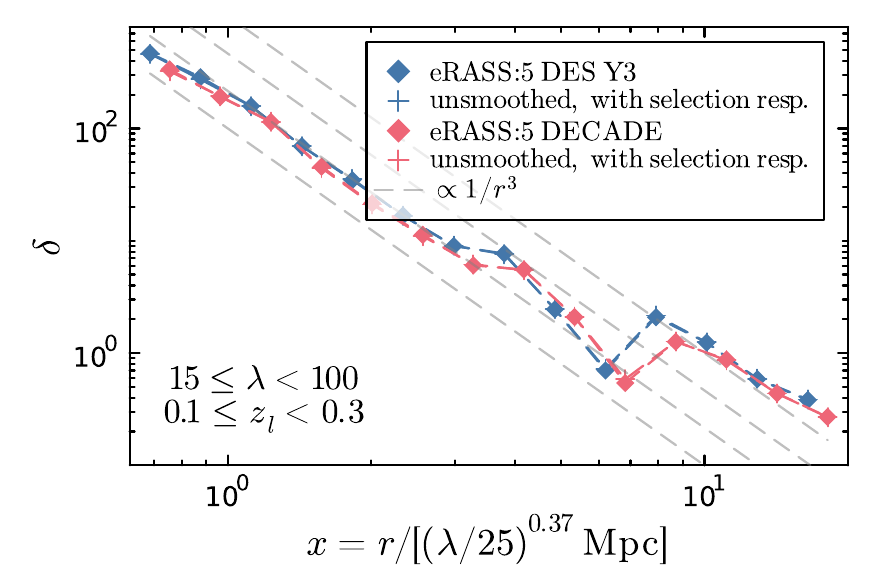}
  \caption{%
   Same as Fig.~\ref{fig:rho_of_x} but additionally showing the result when taking into account the selection response and not smoothing the shear response.
   We calculate the selection response separately for each lens and for each radial bin.
   Error bars are left out for visual clarity.
   We see that our results are barely affected.%
  }
  \label{fig:rho_of_x_selection_response}
\end{figure}

\subsection{Shear and selection response}

In our main analysis, we smooth the shear response on a 2D signal-to-noise and size-ratio grid \citep{Grandis2024,Gatti2021,Anbajagane2025}.
We further do not include the selection response, because we expect it to be only a small effect.
Indeed, for DES Y3 and DECADE, it affects the shear only at the one percent level \citep{Prat2022,Anbajagane2025,Anbajagane2025e}.
Moreover, it is roughly constant as a function of radius and so cannot induce significant radial features like the one we see at $x \sim 6$.
Nevertheless, we have repeated our analysis taking into account the selection response and not smoothing the shear response.
We have calculated the selection response separately for each lens and each radial bin.
This is  to make sure we don't miss any change in the radial shape of our density profiles induced by the selection response.
For simplicity, we do not subtract randoms.
The results are shown in Fig.~\ref{fig:rho_of_x_selection_response}.
As expected, our results are barely affected.

\begin{figure}
  \includegraphics[width=\columnwidth]{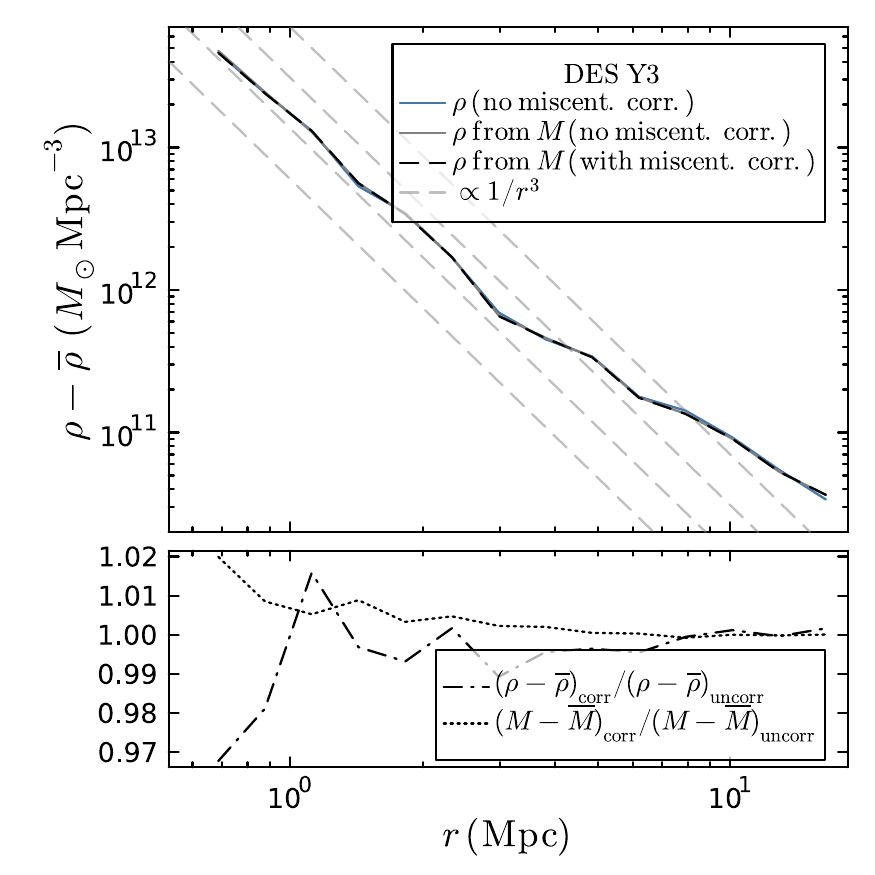}
  \caption{%
   \emph{Top}: Density profiles of the eRASS:5 clusters in DES~Y3 with $\lambda \geq 15$, stacked at fixed physical radius $r$, with and without corrections for miscentering.
   The solid blue line is the baseline result; it is not corrected for miscentering and follows our main analysis (Sect.~\ref{sec:method:shear_to_rho}).
   The solid gray line is the density profile obtained by first obtaining a stacked \emph{mass} profile using Eq.~\eqref{eq:M_from_ESD}, interpolating the result using B-splines, and then taking a derivative of this interpolation.
   There is still no correction for miscentering applied.
   This shows that the density profile calculated from the stacked mass profile matches our original stacked density profile quite well.
   Finally, the dashed black line is the same as the solid gray line, but with a miscentering correction applied (see the main text).
   We see that the effect is small.
   \emph{Bottom}: Ratio of the density profiles with and without the miscentering correction (dash-dotted line).
   Beyond about $1\,\mathrm{Mpc}$, the effect is typically within one percent.
   Physically, we expect a smooth trend.
   Our results show some bumps and wiggles, but this is likely just noise.
   The dotted line shows the same for the mass profiles, where the effect goes into the opposite direction.
  }
  \label{fig:rho_of_r_miscentering}
\end{figure}

\subsection{Miscentering}

Our main analysis does not take into account miscentering, since we expect it to be only a small effect at the large radii we are interested in.
To see this explicitly, Fig.~\ref{fig:rho_of_r_miscentering} shows results that apply a miscentering correction based on the non-parametric miscentering correction formula from \citet{Mistele2024b} to each individual cluster.
For simplicity, we adopt the difference between the BCG and the X-ray center of each cluster as the miscentering offset.
We do not, so far, have an efficient implementation of the miscentering correction formula from \citet{Mistele2024b} in the context the shear to density conversion Eq.~\eqref{eq:rho_from_ESD}.
We do, however, have an efficient implementation in the context of the shear to mass conversion Eq.~\eqref{eq:M_from_ESD} \citep[see Appendix A.2.3 in][]{McGaugh2026}.
Thus, here we compute a density profile by taking the derivative of a stacked mass profile (and dividing by $4 \pi r^2$).
In particular, to keep things simple, we stack in $r$ rather than in $x$ and we calculate the derivative of $M$ from a quadratic B-spline interpolation of the the stacked mass profile.
This simplified procedure suffices to demonstrate that miscentering is unimportant to our results.
Figure~\ref{fig:rho_of_r_miscentering} shows that miscentering has at most a percent-level affect at our results
We also see that the above simplified procedure based on taking a derivative of the stacked mass profile gives results that are close to our more direct density profile measurement, which is a useful cross-check.

\subsection{Cluster member contamination}

We have re-run our analysis without correcting for cluster member contamination and found that this does not significantly affect the results.
Thus, contamination of the source sample from cluster members is not important to our conclusions.
This is as expected since cluster member contamination is not important at the large radii we consider here \citep[e.g.,][]{Grandis2024,Kleinebreil2024,Schrabback2026}.

\begin{figure}
  \includegraphics[width=\columnwidth]{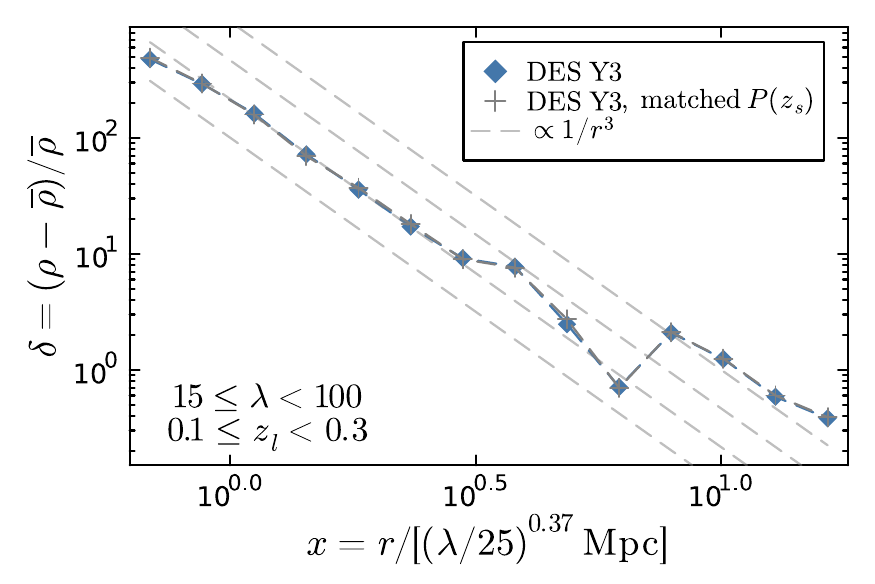}
  \caption{%
   Overdensity profile of the eRASS:5 clusters in DES Y3 using a $P(z_s)$ that is better matched to our weights and lens selection (gray `+' symbols).
   In particular, we use, for each source $s$, the $P(z_s|\hat{c})$ for that source's SOM cell $\hat{c}$ \citep{Myles2021,Grandis2024}.
   The original result is shown as the diamond markers.
   We see that the choice of redshift distribution does not affect our results significantly.
  }
  \label{fig:rho_of_x_per_source_Pzs}
\end{figure}

\subsection{$P(z_s)$ better matched to our analysis}

In our main analysis, we used the global `SOMPZ' $P(z_s)$ redshift distributions provided by DES~Y3 and DECADE for each tomographic bin \citep{Myles2021,Anbajagane2025b}.
These were, however, not calibrated with our choice of weights and our lens selection in mind, which could, in principle, affect the results.
Thus, here we show results where $\langle \Sigma_{\mathrm{crit},ls}^{-1} \rangle$ is instead calculated using the redshift distribution of each individual source's SOM cell, i.e. $P(z_s|\hat{c}_s)$ where $\hat{c}_s$ is the wide SOM cell of source $s$ \citep{Myles2021}.
For simplicity, we do not correct for cluster member contamination and we do not subtract randoms.
Figure~\ref{fig:rho_of_x_per_source_Pzs} shows that this alternative choice of $P(z_s)$ has only a negligible effect on our DES Y3 results.
For DECADE, the $P(z_s|\hat{c})$ distributions are not publicly available, but we expect the effect to be similarly small.

\begin{figure}
  \includegraphics[width=\columnwidth]{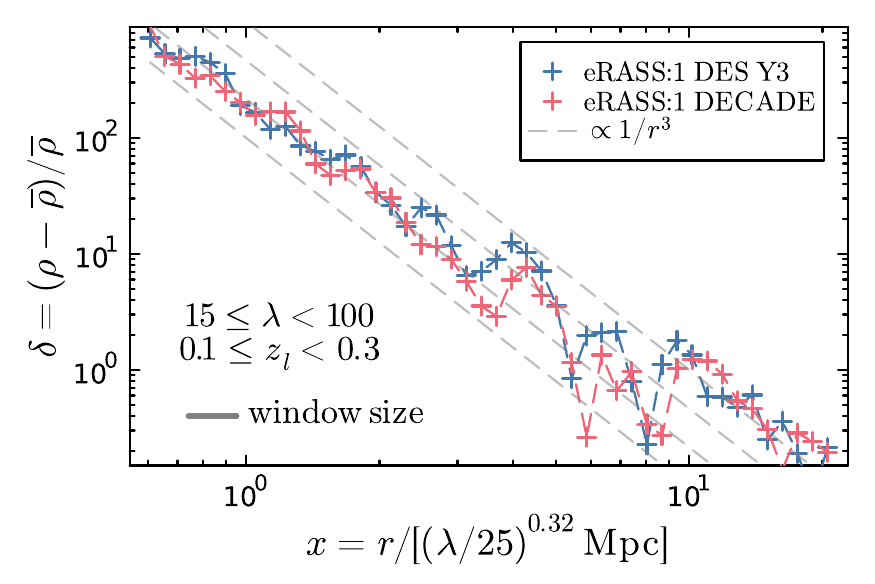}
  \caption{%
   Overdensity profiles of the eRASS1 cosmology clusters within the DES Y3 (blue) and DECADE (red) footprints, using overlapping sliding windows.
   We use a different rescaling power $p = 0.32$ than for eRASS:5.
   We still see the $x \sim 6$ feature, though not as clearly as for eRASS:5.
   We can recover the eRASS:5 result (with $p = 0.37$ and a strong feature at $x \sim 6$), if we cross-match the eRASS1 and eRASS:5 clusters and replace the richness measurements in the eRASS1 catalog with those from the eRASS:5 catalog.
  }
  \label{fig:rho_of_x_erass1}
\end{figure}

\subsection{eRASS1}

As discussed in Sect.~\ref{sec:results:checks}, the $x \sim 6$ feature seems to be most pronounced at high richness and low redshift.
Many of these clusters were already part of the eRASS1 cluster catalog \citep{Bulbul2024}.
Thus, we should see a similar feature when replacing the eRASS:5 catalog with the eRASS1 catalog.
To test this, Fig.~\ref{fig:rho_of_x_erass1} shows results using the eRASS1 cosmology sample.
We indeed see a similar feature around $x \sim 6$, although it's less pronounced and the feature is most consistent between DES and DECADE for a different power $p$ in our radial rescaling procedure, namely $p=0.32$.
There are also again possible hints of a secondary feature around $x = 3$.
These small differences to eRASS:5 may be due to differences in the richness assigned to the clusters.
Indeed, we have re-run our analysis using the eRASS1 cosmology sample, but with the richness measurements from eRASS:5 (after cross-matching based on coordinates and redshift and restricting the sample to only those clusters for which we find a match).
With the eRASS1 richness replaced by the eRASS:5 richness, we find results very close to those for eRASS:5, including the power of $0.37$ being preferred and the feature being more pronounced.
We attribute this to the improved richness measurements for eRASS:5 \citep{erass5catalog}.

\subsection{Shear vs reduced shear}

When converting shear to density we neglect the difference between shear and reduced shear (Sect.~\ref{sec:method:shear_to_rho}).
We expect this difference to be unimportant at the large radii we consider since shear and reduced shear differ only by a factor of $1 - \kappa = 1 - \Sigma/\Sigma_\mathrm{crit}$, which is close to one at large radii.
Indeed, we have re-run our analysis taking into account the difference between shear and reduced shear using the formulas from \citet{Mistele2024b} and we have found only barely perceptible differences.
For practical reasons, this conversion works only with linear interpolation, so we used linear interpolation for this test.\footnote{
   The reason is that our code for handling the non-linear conversion from reduced shear to $\Delta \Sigma$ does not work with values of $G \cdot \Sigma_\mathrm{crit}^{-1}$ larger than one.
   Such values, however, occur for some individual clusters when using quadratic interpolation.
}
The conversion from reduced shear to shear becomes non-linear when $\kappa$ is significant.
Thus, in principle, the noise may no longer cancel out as nicely and the results may be biased.
The above test shows that, in practice, this does not affect our results.

\begin{figure}
  \centering
  \includegraphics[width=\columnwidth]{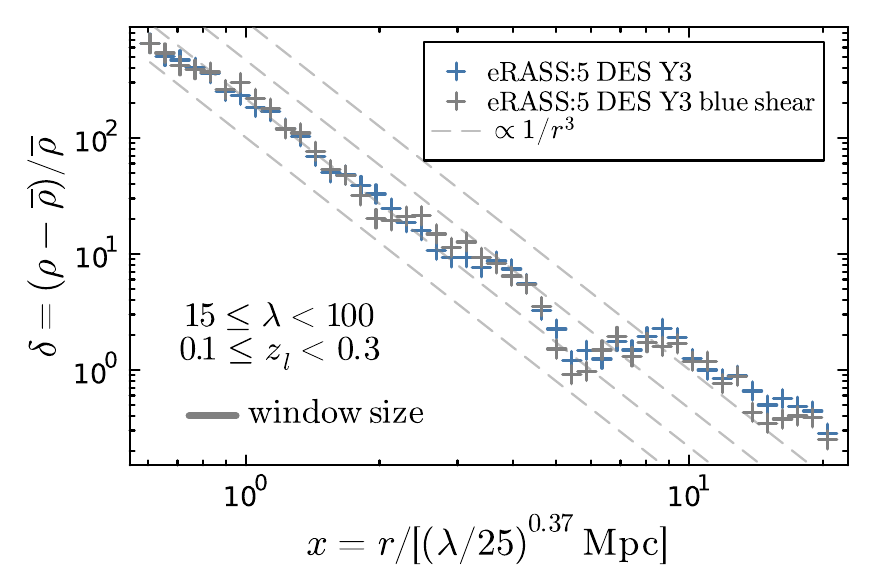}
  \caption{%
   Overdensity profiles obtain using the ``Blue shear'' subset of sources from DES~Y3 (gray markers) from \citet{McCullough2024}.
   The result using the full DES~Y3 source sample is shown as blue markers.
   For simplicity, we do not perform any subtraction of randoms or correction for cluster member contamination in both cases.
   We see that restricting to the blue shear sources leaves our result almost unchanged.
   This indicates that intrinsic alignment, which is expected to be less strong for blue sources, does not affect our results.
  }
  \label{fig:blue_shear}
\end{figure}

\subsection{Intrinsic alignment}

Intrinsic alignment is expected to be much stronger in red than in blue galaxies.
Thus, we can test for the effect of intrinsic alignment by restricting our analysis to blue source galaxies.
In particular, Fig.~\ref{fig:blue_shear} shows results when restricting to the ``Blue shear'' subset of source galaxies of DES~Y3 \citep{McCullough2024}.
For simplicity, Fig.~\ref{fig:blue_shear} does not correct for cluster member contamination and does not subtract randoms.
Our results do not change significantly, implying that our results are not significantly affected by intrinsic alignment.

\begin{figure}
  \includegraphics[width=\columnwidth]{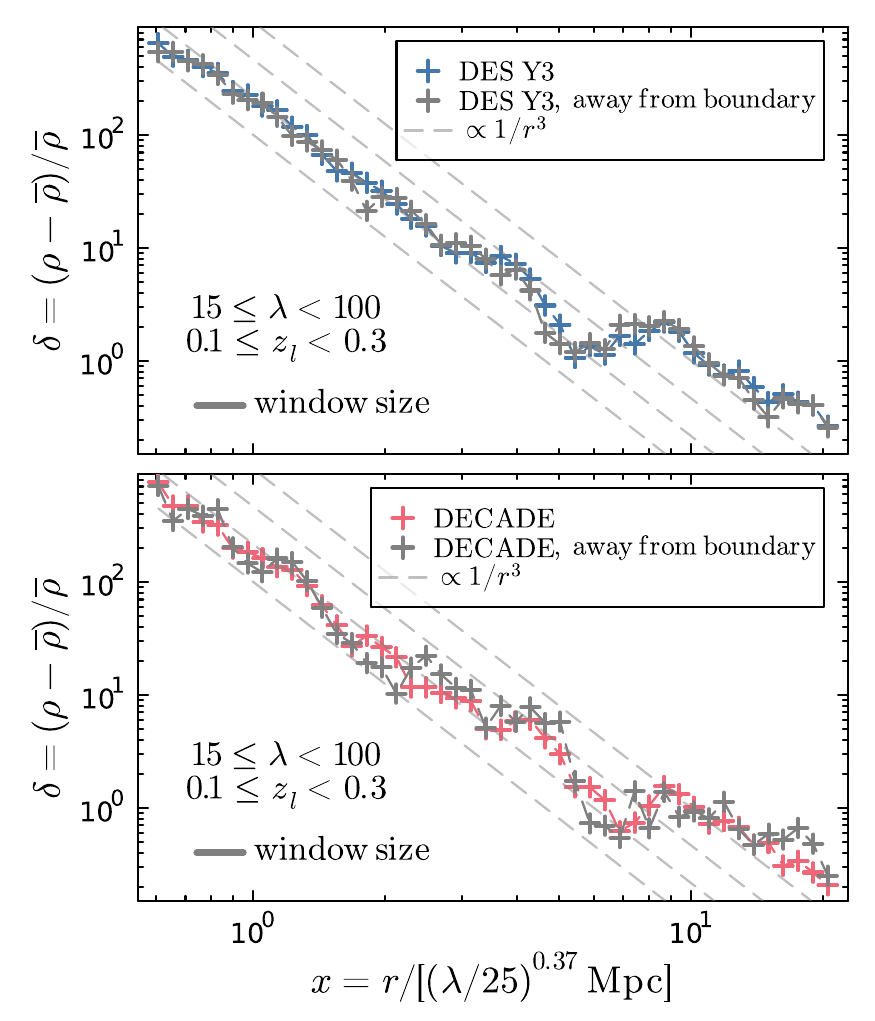}
  \caption{%
   Overdensity profiles of the eRASS:5 clusters, with all clusters within $25\,\mathrm{Mpc}$ of the lens survey edge removed (gray markers).
   The original results including all clusters are shown as colored markers.
   We see that the feature around $x \sim 6$ remains strong, indicating that the feature is not due to a lens survey edge effect.%
  }
  \label{fig:rho_of_x_survey_edge}
\end{figure}

\subsection{Lens survey edge}

Clusters located near the edge of the DES and DECADE footprints may have only partial weak-lensing data.
In such cases, the clusters are assigned smaller weights in the relevant radial range, due to the partially missing data there.
This may, in principle, introduce a spurious radial trend in our results, if the clusters affected are not representative of the sample as a whole.
However, as we have seen above, our results do not change significantly if we use the same weights in all radial bins (Fig.~\ref{fig:rho_of_x_weights}).
Thus, such effects on the weights are likely insignificant in practice.
Still, an incomplete azimuthal average due to the lens survey edge might have some effect.
We test for such effects by re-running our analysis with all clusters within $35\,\mathrm{Mpc}$ of the lens survey edge removed, see Fig.~\ref{fig:rho_of_x_survey_edge}.
The smaller sample leads to increased noise, but we still clearly see the $x \sim 6$ feature.
This shows that our feature is not a lens survey edge effect.

\subsection{Scatter in mass proxy}

We use $(\lambda/25)^{0.37}$ as a proxy for cluster size.
If that is indeed a reasonable proxy, our results should tolerate some amount of scatter in this quantity.
We have therefore checked how our results change when artificially adding scatter to $\lambda$.
We find that our results are mostly unchanged when adding $0.14\,\mathrm{dex}$ (roughly $35\%$) scatter to $\lambda$ and our results start to look noisy around $0.19\,\mathrm{dex}$ (roughly $55\%$) scatter.
This confirms that we can tolerate a reasonable amount of scatter.

\section{Significance testing procedure}
\label{sec:appendix:significance}

Here, we describe the details of our significance testing procedure from Sect.~\ref{sec:results:significance}.
We begin by fitting broken power laws to our stacked density profiles for DES~Y3 and DECADE by minimizing the $\chi^2$, taking into account the full covariance matrix.
We consider two different models: One is a broken power law
\begin{align}
 \delta_{\mathrm{broken}}(r) = \delta_0 \cdot \begin{cases}
  \left(\frac{r_0}{r}\right)^3, & \mathrm{if }\; r < r_0 \\
  \left(\frac{r_0}{r}\right)^\alpha , & \mathrm{else} 
 \end{cases}
\end{align}
with the normalization $\delta_0$, the transition radius $r_0$, and the outer slope $\alpha < 3$ being free parameters.
The second model replaces the sharp transition at $r_0$ with a smooth transition
\begin{align}
 \delta_{\mathrm{smoothed}}(r) = \delta_0 \left[\left(\frac{r_0}{r}\right)^3 + \left(\frac{r_0}{r}\right)^\alpha \right] \,.
\end{align}
Since these do not capture the slight trend to a shallower slope than $-3$ at $x \lesssim 1$, we consider only the $x > 1$ region for these broken power law models.
For the \citet{Diemer2023} fits, we do not exclude the $x < 1$ region.
We then generate, separately for DES and DECADE, $N = 50\mathord,000\mathord,000$ samples following the best-fit model and following the full observational covariance matrix.
For each sample, we check if there is a feature at least as strong as in our observed results.
Finally, we calculate the frequencies $N_\mathrm{feature}/N$, where $N_\mathrm{feature}$ is the number of samples that exhibit a feature at least as strong as in our observed results.

For the observational results we use $p=0.37$ and apply the cuts $15 \leq \lambda < 100$ and $0.1 \leq z_l < 0.3$.
We use sliding windows with the jackknife covariance matrix.
Using sliding windows avoids complications due to the fact that, above, we used slightly different radial bins for DES and DECADE.
This leads to neighboring data points being highly correlated, which is taken into account by the jackknife covariance matrix.
Using sliding windows also leads to a more reliable measure of the width of the feature.
We judge whether a sample has a feature at least as strong as our observed data in the following way:
For each sample and separately for DES and DECADE, we first find the radial range over which the largest uptick relative to the statistical uncertainty occurs.
For simplicity, we take the statistical uncertainty to be the mean between that of the start and end of the uptick's radial range.
We then check, for each sample, whether all of the following conditions are satisfied; these conditions reflect the salient features in our observed result:
\begin{enumerate}
\item The DES and DECADE samples agree with each other at least as well as they agree in the observed data, as quantified by the $\chi^2$\footnote{
 Calculating $\chi^2$ requires inverting a noisy jackknife covariance matrix.
 We therefore should multiply our $\chi^2$ by the Hartlap factor \citep{Hartlap2007}.
 However, since we use this $\chi^2$ only to compare observed data and samples and since we use the same covariance matrix for both, any overall multiplicative factor is inconsequential and can be omitted.
}.
\item For both DES and DECADE the uptick is at least as large as in the observed data (relative to the level of uncertainty, see above).
\item The radial ranges of the largest upticks in the DES and DECADE samples overlap.
\item For both DES and DECADE, the radial range of the largest uptick is at least as extended as in the observed data (in $\log x$).
\item For both DES and DECADE and outside the uptick's radial range, the samples agree with the best-fit model at least as well as the observed data. Here ``outside the uptick's radial range'' means that we cut out the uptick's radial range and an equally large (in $\log x$) region immediately before this uptick (where typically there is a corresponding downturn).
 The degree of agreement is quantified by the reduced $\chi^2$ (reduced because in general we remove a different radial range in the observed data and in the samples).
\end{enumerate}

\section{Circular velocity}
\label{sec:appendix:MOND}

\begin{figure}
  \includegraphics[width=\columnwidth]{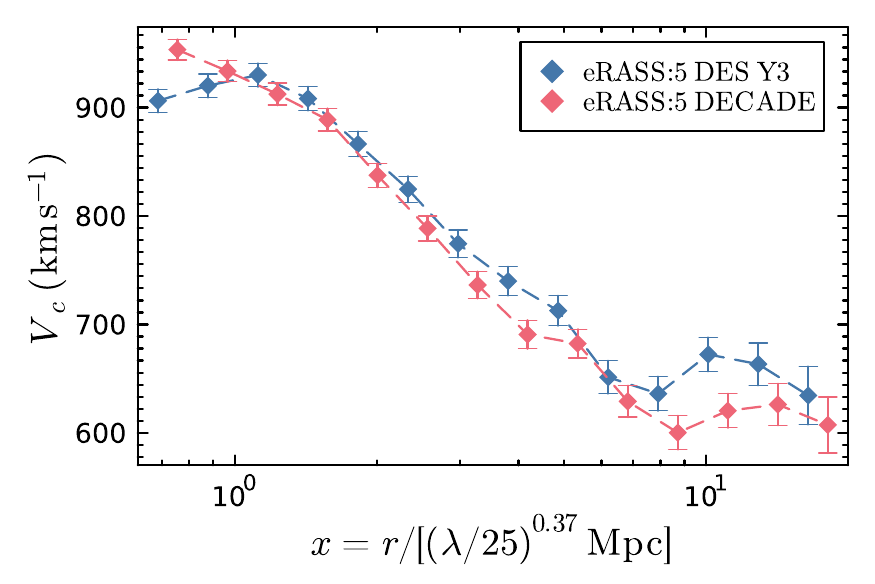}
  \caption{%
   The stacked circular velocity profiles $V_c (r)$ of the eRASS:5 clusters.
   In particular, we calculate a stacked $V_c^2 = G_N M(r)/r$ profile and then take the square root.
   This relation between the squared circular velocity and the mass profile holds in spherical symmetry and we adopt it here for simplicity.
   We see a clear decline in the radial region where our stacked density profiles decline as $r^{-3}$.
  }
  \label{fig:Vc_of_x}
\end{figure}

As a relevant additional result in the context of MOND, Fig.~\ref{fig:Vc_of_x} shows the circular velocities $V_c = \sqrt{G_N M(r) / r}$, which may be a more intuitive quantity in that context.
Here, we take ``$M(r)$'' to not include the background's mass $(4\pi/3) \overline{\rho} r^3$, since that does not contribute to the velocities of test particles.
The $r^{-3}$ behavior of the density leads to a decline in $V_c$.
This contrasts with the galactic circular velocities which the weak-lensing analysis of \citet{Mistele2024} found to be flat, though at smaller radial scales of hundreds of $\mathrm{kpc}$.

\end{appendix}

\end{document}